\documentclass[aps,twocolumn,pra,superscriptaddress,amsmath,showpacs,tightenlines]{revtex4-1}
\usepackage{bbm}
\usepackage{}
\usepackage{amssymb}
\usepackage{amsmath}
\usepackage{graphicx}
\usepackage{subfigure}
\usepackage{natbib}
\usepackage{epsfig}
\usepackage{amsfonts}
\usepackage{mathrsfs}
\usepackage{xcolor}
\usepackage[toc,page,title,titletoc,header]{appendix}

\begin{document}

\title{Controllable single-photon wave packet scattering in two-dimensional resonator
array by a giant atom}
\author{Weijun Cheng}
\affiliation{School of integrated circuits, Tsinghua University, Beijing 100084, China}
\author{Zhihai Wang}
\email{wangzh761@nenu.edu.cn}
\affiliation{Center for Quantum Sciences and School of Physics, Northeast Normal University, Changchun 130024, China}
\author{Yu-xi Liu}
\email{yuxiliu@mail.tsinghua.edu.cn}
\affiliation{School of integrated circuits, Tsinghua University, Beijing 100084, China}
\affiliation{Frontier Science Center for Quantum Information, Beijing, China}

\begin{abstract}
Nonlocal interactions between photonic resonator array and giant atoms have attracted extensive attentions. Optimization and control of quantum states via giant atoms have been shown. We here study the dynamical scattering of a single-photon wave packet by a giant atom coupled to a two-dimensional photonic resonator array via multiple spatial points. Using several iterations of time evolutions, we can prepare an expected wave packet with a stable size and use it as the incident state for the scattering process. We show that spatially symmetric or asymmetric target scattering states of single-photon wave packet can be generated by adjusting the coupling strengths between the giant atom and different lattice sites of the resonator array. Furthermore,  the dynamical scattering of the wave packets enables us to  study  the atomic excitation and  propagating properties of the scattering states. We find that the atomic excitation has negligibly small probability during the scattering process.  Our study may provide a new way to generate an expected photon state via photon scattering by a giant atom in two-dimensional photonic array.
\end{abstract}

\maketitle

\section{introduction}
Single-photon sources have important applications in quantum optics, quantum communication~\cite{Couteau,Luo}, quantum metrology~\cite{Eisaman,Chunnilall}, and quantum computation~\cite{Maring}. The best-performing single-photon source is based on the conditional detection for correlated photon pairs generated by spontaneous parametric down-conversion process~\cite{DP}. Single photons source can also be obtained via linear optics~\cite{Darquie},  spontaneous emission  of quantum emitters~\cite{two-level1,two-level2,two-level3,two-level4} and other approaches~\cite{Hofmann,Yuan}.  In practice, the single-photon source produces a light pulse with no more than one photon and is characterized by the second-order correlation function~\cite{Senellart,Zhai}.  A single-photon pulse is usually described by a single-photon wave packet state, which is a superposition of single-photon Fock states of infinite plane wave modes.
To enhance the collection efficiency and the spontaneous emission rate of single-photons, the quantum emitters  for single-photon sources are usually coupled to one-dimensional waveguides~\cite{Zumofen,Chang,YChen}  or single-mode resonators~\cite{Peng-NC}. Moreover, to realize quantum information transfer via single-photons, arbitrarily controllable single-photon wave packets are highly desired~\cite{Cai,Srivathsan,Tian}. For example, specifically temporal shaped single-photon wave packets can be  used for optical amplifiers~\cite{Rephaeli2,Linares,FWSun}, deterministic quantum state transfers~\cite{Cirac}, and implementation of quantum logic gates~\cite{Heuck}.

Photon scattering by atoms or artificial atoms has emerged as one of powerful and straightforward quantum optical tools for manipulating the single-photon wave packets in the waveguide~\cite{POGuimond,Shi,Shen,Shi2,Xu,Zheng,Zhou,Nie}. In these studies~\cite{POGuimond,Shi,Shen,Shi2,Xu,Zheng,Zhou,Nie}, the atoms are considered as point particles and are assumed to be coupled to the waveguide via a spatial point due to its negligible size compared to the wavelengths of the modes in the waveguide. In a seminal work, the superconducting qubits~\cite{Gu-PR}, acting as giant artificial atoms, were designed to interact with surface acoustic waves via multiple spatial points~\cite{Andersson}. From then on, the interactions between superconducting giant atoms and surface acoustic wave waveguides or resonators were extensively studied~\cite{SAW1,SAW2, SAW3,SAW4,SAW5,Xin2}. The  giant atoms were also extended from superconducting quantum systems to cold atom systems~\cite{GonzalezTudela} and ferromagnetic spin systems~\cite{ZQWang}. Recently, the interactions between the giant atom and microwave waveguides or resonators were also studied~\cite{Kannan,xin2} and various phenomena have been shown. For example,  retardation effect~\cite{Cheng} in the waveguide can be found even for a single giant atom in contrast to many point-like atoms. This phenomenon is resulted from the interference of photons between the multiple spatial coupling  points of a single giant atom to the waveguide. The retardation effect in the waveguide coupled to many point-like atoms is from the interference of photons between atoms. The multiple coupling points can also result in nonexponential decay~\cite{Andersson2,Guo} even the coupling between the giant atom and the waveguide is not strong. It was also shown that the decoherence-free for all states of many two-level giant atoms can be achieved for a certain arrangement of the coupling points of the giant atoms to the waveguide~\cite{Kockum2}. However, the decoherence free for many point-like atoms can be achieved only for special quantum states~\cite{wu1,wu2}. In particular, multiple coupling points can give rise frequency-dependent relaxation rates and Lamb shifts~\cite{Kockum}, which cannot occur for the system of the point-like atoms. Moreover, single-photon scattering has also been studied by coupling giant atoms to a one-dimensional waveguide via multiple spatial points~\cite{Chen,Peng,NLiu,WZhao}.

Two-dimensional  waveguide or coupled-resonator array can be used to simulate quantum phase transitions of light~\cite{Greentree}, realize quantum walks of correlated photons~\cite{TD1,Jiao}, photosynthetic energy transport~\cite{Mohseni}. The theoretical study shows that the relaxing of the giant atom can be avoided when the giant atom interacts with an environment of two-dimensional  resonator array~\cite{Ingelsten}.  Moreover, coherent interaction between a quantum emitter and the edge states in two-dimensional optical topological insulators has been studied~\cite{QE-T1}. Topological multimode waveguide QED with two-dimensional topological photonic system coupled to a quantum emitter has also been demonstrated~\cite{QE-T2}. Furthermore, collective phenomenon in the multiscattering of a single photon has been shown in the system of many atoms coupled to a two-dimensional resonator array~\cite{DZXu}.  However, the single photon scattering by the giant atom coupled to  a two-dimensional resonator array is not studied to our best knowledge.

Here, motivated by previous studies on single-photon scattering by a giant atom in one-dimensional resonator array or many alined atoms in two-dimensional photonic resonator arrays~\cite{DZXu}, we study single-photon scattering by a giant atom coupled to two-dimensional photonic resonator arrays via multiple spatial points.
Due to the interference of photons between discrete coupling points, the photon distribution in the two-dimensional resonator array can be engineered by varying the coupling strengths of the coupling points of the giant atom to the array or the distance between the coupling points. Thus, controllable two-dimensional photonic devices can be obtained.
In our study, we first show the dispersion effect on the size of the propagating wave packet and show how a wave packet with a stable size can be obtained.
We proceed to analyze the dynamical scattering of single-photon wave packets by either a small or a giant atom, based on the scattering matrix.
By adjusting the coupling strengths of different coupling points of the giant atom to  two-dimensional photonic resonator array, we find that the symmetric or asymmetric photon distribution of the scattered single-photon wavepacket can be optimized. In addition, we also explore the propagating properties of the incident and target scattering wave packets via  the propagating fidelity and study the excitation of giant atoms by single-photon wave packet. Finally, we compare the numerical results with analytical ones and show the effectiveness of our analytical studies.

The paper is organized as follows. In Sec.~\ref{Theoretical}, we describe the theoretical Hamiltonian model and derive a formula of the scattering matrix. In Sec.~\ref{Preparation of Nonspreading Wave Packet}, we introduce a theory to produce the desired incident single-photon wave packet in the two-dimensional photonic resonator array. The dispersion effect on the propagation of the single-photon wave packet is also studied. In Sec.~\ref{Propagation and Scattering of Wave Packet}, we use the scattering matrix to study the dynamical scattering of single-photon wave packet by a small atom. In Sec.~\ref{Controllable}, we derive an optimization function and study controllable single-photon scattering based on the giant atom. In Sec.~\ref{Chiral}, we study single-photon scattering by a giant atom coupled to 2D resonator array via 2D spatial points. In Sec.~\ref{conclusion}, we summarize our results and analyze the feasibility of the experiment.

\section{Theoretical  Model and  Scattering Matrix} \label{Theoretical}

\subsection{Theoretical Model and Hamiltonian}

\begin{figure}[htbp]
\centering
  \includegraphics[width=8cm]{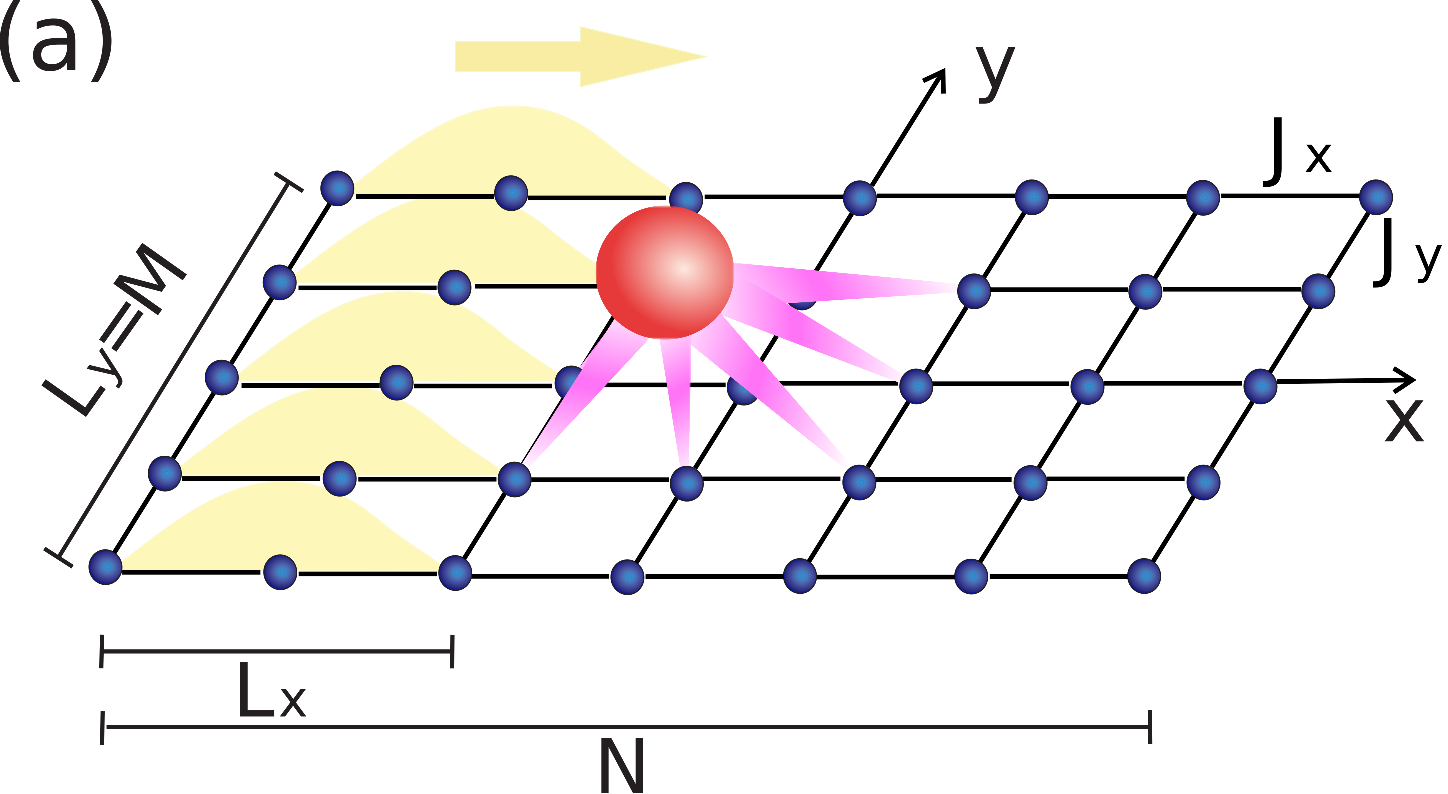}\nonumber\\
 \includegraphics[width=8cm]{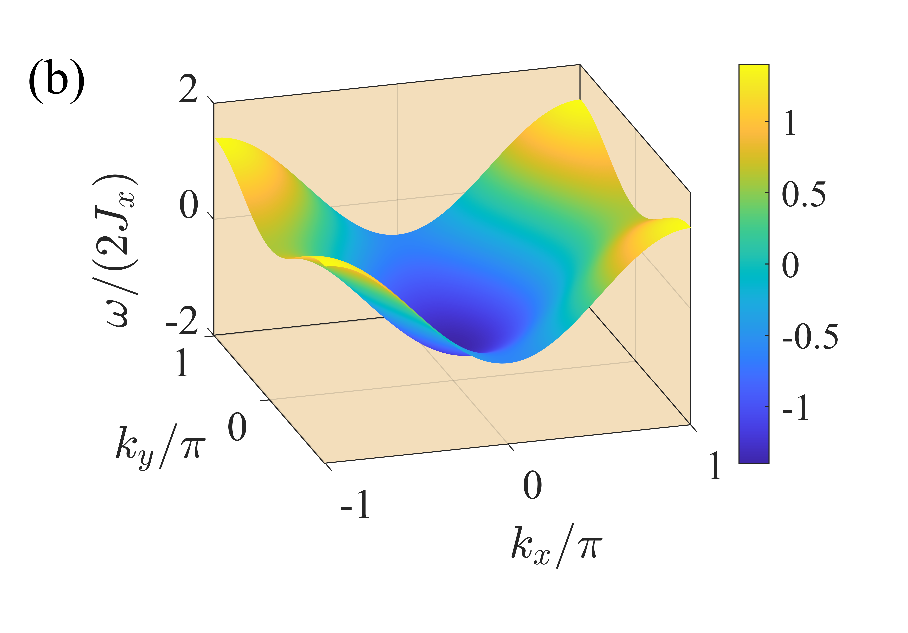}
\caption{(a) Schematic diagram for the 2D photonic resonator array coupled to a two-level giant atom (red ball) via multiple spatial points (purple points). (b) The energy spectrum $\omega(\vec{k})$ of the 2D photonic resonator array. The parameters are set as: $\omega_l=0$, $J_y/(2J_x)=0.2$.}
\label{2DEnergyspectrum}
\end{figure}

As schematically shown in Fig.~\ref{2DEnergyspectrum}(a), we study a system that a two-level giant atom,
with the transition frequency $\omega_a$ between the ground state $|g\rangle$ and excited state $|e\rangle$, is coupled to a two-dimensional (2D) photonic resonator array via multiple spatial points. The Hamiltonian of the whole system can be written as  \begin{eqnarray}
H=H_0+H_{{\rm G,I}}.
\label{H}
\end{eqnarray}
Where the free Hamiltonian $H_0$ of the 2D photonic resonator array and the giant atom is given by
\begin{equation}
H_{0}=H_{0}^{\prime}+\omega_a|e\rangle\langle e|,
\label{H0}
\end{equation}
with the Hamiltonian $H^{\prime}_0$ of the 2D photonic resonator array
\begin{eqnarray}
\begin{split}
H_{0}^{\prime}=&\sum_{m=-M}^M\sum_{n=-N}^N \left\{\omega_la_{n,m}^{\dag}a_{n,m}\right. -(J_x a_{n,m}^{\dag}a_{n+1,m}\\
 &+\left.J_y a_{n,m}^{\dag}a_{n,m+1}+{\rm H.c})\right\},
 \end{split}
\label{2DHr}
\end{eqnarray}
hereafter, we assume $\hbar=1$.  $a_{n,m}$ and $a_{n,m}^{\dag}$ are the annihilation and creation operators of the photonic mode on the lattice site $(n,m)$, respectively.  The integer numbers $n$ and $m$ denote the positions in $x$-direction and $y$-direction, respectively.   We also assume that all photonic modes have the same frequency $\omega_l$. Moreover, we assume that there are only  the nearest-neighbor hopping interactions between different photonic modes. The parameters $J_y$ and $J_x$ denote the hopping strengths along $y$-axis and $x$-axis, respectively.
 $\sigma_{+}=|e\rangle\langle g|$ ($\sigma_{-}=|g\rangle\langle e|$) is the raising (lowering) operator of the two-level  giant atom, which locates at the lattice site $(0,0)$. The interaction Hamiltonian $H_{{\rm G,I}}$ between the giant atom and the photonic resonator array via multiple coupling points is given as
 \begin{eqnarray}
H_{{\rm G,I}}=\sum_{m_1=-M_1}^{M_1}\sum_{n_1=-N_1}^{N_1}\frac{g}{\Lambda}\left\{ \lambda_{n_1,m_1}a_{n_1,m_1}\sigma^{+}+{\rm H.c}\right\}.\nonumber \\
\label{HGIr}
\end{eqnarray}
The summation is taken for all coupling points $(n_{1},m_{1})$.
To quantify the coupling between the giant atom and the 2D resonator array under different configurations, we define an overall coupling strength $g$. The relative coupling strengths between the giant atom and resonator array at different coupling sites are characterized by the parameter $\lambda_{n_1,m_1}$, which is normalized  by the parameter  $\Lambda=\sum_{n_1,m_1} |\lambda_{n_1,m_1}|$. That is,
the parameter $g\lambda_{n_1,m_1}/\Lambda$ represents the coupling constant between the giant atom and the photonic mode at a specific lattice site $(n_1,m_1)$. For convenience and without loss of generality, we assume that the site $(0,0)$ is the center of all coupling sites and the total number of the coupling points is $(2N_1+1)(2M_1+1)$.

 The free Hamiltonian $H_{0}^{\prime}$ in Eq.~(\ref{2DHr}) of the photonic resonator array with the total lattice numbers $2N+1$ and $2M+1$ in the $x$-axis and $y$-axis can be rewritten as
\begin{eqnarray}
H_{0}^{\prime}(\vec{k})=\sum_{k_x,k_y} \omega(\vec{k}) a_{k_x,k_y}^{\dag}a_{k_x,k_y},
\label{2DH}
\end{eqnarray}
 in the momentum space via the Fourier transform
 \begin{equation}\label{F3}
a_{n,m}= \frac{1}{\sqrt{(2M+1)(2N+1)}}\sum_{k_x,k_y}a_{k_x,k_y}e^{ik_xn+ik_ym}
\end{equation}
 under the periodical boundary condition. Here, the frequency
\begin{eqnarray}
\omega(\vec{k})\equiv\omega(k_{x},k_{y})=\omega_l-2J_x\cos(k_x)-2J_y\cos(k_y)
\label{ES}
\end{eqnarray}
has a cosine-type nonlinear structure. The wavenumbers $k_x$ and $k_y$ are defined as $k_x \in\{-\pi, -\pi(N-1)/N, \cdots, 0, \cdots, \pi(N-1)/N, \pi\}$ and $k_y \in \{-\pi, -\pi(M-1)/M, \cdots, 0, \cdots, \pi(M-1)/M, \pi \}$. Equation (\ref{ES}) clearly shows that the energy spectrum  $\omega(\vec{k})$ is in the range $ [\omega_l-2J_x-2J_y,\omega_l+2J_x+2J_y]$ and is an even function of $k_x$ and $k_y$. Thus, as shown in Fig.~\ref{2DEnergyspectrum}(b),  $\omega(\vec{k})$ exhibits a symmetric characteristics in both $k_x$ and $k_y$ directions.

In the momentum space under the Fourier transform in Eq.~(\ref{F3}), the interaction Hamiltonian $H_{{\rm G,I}}$ in Eq.~(\ref{HGIr}) can be rewritten  as
\begin{eqnarray}
H_{{\rm G,I}}(\vec{k})=
\sum_{k_x,k_y}\left\{G(\vec{k})a_{k_x,k_y}\sigma^{+}+{\rm H.c}\right\}.
\label{HGIk}
\end{eqnarray}
 The coupling constant $G(\vec{k})$ between the giant atom and $\vec{k}=(k_{x}, k_{y})$ mode photon is
\begin{eqnarray}
G(\vec{k})=\frac{g}{\Lambda}\sum_{n_1,m_1}\lambda_{n_1,m_1}e^{ik_xn_1+ik_ym_1},
\label{Gk}
\end{eqnarray}
which is derived from the summation of the coupling strengths of different coupling points in the position space and characterizes the interference induced by the different spatial points.

\subsection{Scattering Matrix}
For the sake of completeness  and convenience of following study, we now briefly summarize the theoretical description of the scattering process.  In the scattering theory, the relation between final and initial states of the system can be expressed by the scattering matrix,  which can be obtained via the scattering operator $S=U(t_i,t_f)$ from the initial time $t_i=-\infty$ to the final time $t_f=+\infty$ as~\cite{POGuimond,Taylor}
 \begin{eqnarray}
S=1-2\pi i\delta(E_f-E_i)T(E_i),
\label{Smatrix}
\end{eqnarray}
with
\begin{eqnarray}
T(E_i)=\sum_{n=0}^{\infty}H_{{\rm G,I}}\left(\frac{1}{E_i-H_{0}+i0^+}H_{{\rm G,I}}\right)^n.
\label{Tmatrix0}
\end{eqnarray}
Here, the subscripts $i$ and $f$ denote initial and final, respectively. In our study, the number of excitations is conserved since the Hamiltonian is derived in the rotating wave approximation. That is,  the energy is conserved in the scattering process, which is guaranteed by $\delta$-function. For single-photon scattering, $E_i$ and $E_f$ correspond to the energies of a single-photon in the initial state $|\vec{k}_i,g\rangle=|\vec{k}_i\rangle\otimes|g\rangle$ with the photon wave vector $\vec{k}_i$  and the final state $|\vec{k}_f,g\rangle=|\vec{k}_f\rangle\otimes|g\rangle$ with the photon wave vector $\vec{k}_f$, which satisfy equations $H_0|k_i\rangle=E_i|k_i\rangle$ and $H_0|k_f\rangle=E_f|k_f\rangle$, respectively. Here, the energies of the single-photons corresponding to the initial and final states are given as

\begin{equation}
E_{\alpha}=\omega(\vec{k}_{\alpha})=\omega_l-2J_x\cos(k_{\alpha, x})-2J_y\cos(k_{\alpha, y}),
\end{equation}
for $\vec{k}_{\alpha}=(\vec{k}_{\alpha,x}, \vec{k}_{\alpha,y})$ with $\alpha=i$ or $\alpha=f$.
We note that the first term in Eq.~\eqref{Smatrix} describes the free evolution of the single-photon packet, while the operator $T(E_i)$ describes the scattering effect of the atoms on the single-photons. According to Ref.~\cite{POGuimond}, the matrix elements of the scattering operator in Eq.~\eqref{Smatrix} can be written as (see details in Appendix~\ref{Appendix})
\begin{widetext}
 \begin{eqnarray}
\langle g, \vec{k}_f|S|\vec{k}_i,g\rangle
&=&\langle \vec{k}_f|\vec{k}_i\rangle-\frac{i}{2\pi}\sum_{q}\frac{\delta(\vec{k}_f-\vec{q})}{|\nabla E_f|_{\vec{k}_f=\vec{q}}}\frac{G^{\ast}(\vec{k}_f) G(\vec{k}_i)}{E_i-\omega_a-\Sigma+i0^+},
\label{Tmatrix}
\end{eqnarray}
\end{widetext}
with the wave vector $\vec{q}$, which is a value of $\vec{k}_f$ satisfying $E(\vec{k}_f)-E(\vec{k}_i)=0$ for a given $\vec{k}_i$ of the incident single-photon wavepacket. The  self-energy parameter $\Sigma$ of the giant atom is given as
 \begin{eqnarray}
\Sigma&=&\left\langle e,0\left|\frac{H_{G,I}}{E_i-H_{0}+i0^+}H_{G,I}\right|0,e\right\rangle\nonumber \\
&=&\frac{1}{4\pi^2}\int dk_xdk_y\frac{|G(\vec{k})|^2}{E_i-\omega(\vec{k})+i0^+}, \nonumber \\
\label{self-energy}
\end{eqnarray}
which is interpreted as the interaction energy between the giant atom and the photonic resonator array.  The  imaginary part of $\Sigma$ corresponds to the decay of an atomic excitation via photons and the real part of $\Sigma$ corresponds to the Lamb shift. From Eq.~(\ref{Tmatrix}), we can obtain the scattering probability $s=|\langle g,\vec{k}_i|S|\vec{k}_i,g\rangle|^2$ as
 \begin{eqnarray}
s=\left|1-\frac{i|G(\vec{k}_i)|^2}{2\pi|\nabla E_f|_{\vec{k}_f=\vec{k}_{i}}(E_i-\omega_a-\Sigma+i0^+)}\right|^2.
 \label{TP}
\end{eqnarray}
when we only consider  $\vec{q}=\vec{k}_{i}$.

\section{Dispersion effect on propagation of single-photon wave packet in 2D photonic resonator array}\label{Preparation of Nonspreading Wave Packet}

Our goal is to study the scattering of single-photon, which  is described  by  a single-photon wave packet state.  We know that a single-photon wave packet state can usually be expressed as~\cite{YWang}
\begin{eqnarray}
|\psi(\vec{k}_c)\rangle=\sum_{\vec{r}} \phi(\vec{r})e^{i\vec{k}_c\cdot\vec{r}} a_{\vec{r}}^{\dag}|0\rangle,
\label{wavepacket}
\end{eqnarray}
where $a_{\vec{r}}^{\dag}$ is the creation operator at the position $\vec{r}$ and
$\phi(\vec{r})$ is the spatial distribution function of the single-photon wave packet with $\sum_{\vec{r}}|\phi(\vec{r})|^2=1$. $\vec{k}_c$
 corresponds to the wave vector of the wave packet center. However, such a wave packet cannot stably propagate due to the dispersion even that the wave packet is not scattered by other subjects.
A wave packet that maintains its shape and amplitude over propagation time within the resonator array is of great significance for the analysis of scattering behavior. The issues related to wave packets have been extensively studied in previous research.
 However, for completeness of the paper and comparison with the scattered single-photon wave packet, we analyze the dispersion of the single-photon wave packet in a two-dimensional resonator array when the single-photon wave packet propagates with a given group velocity.
 We first summarize and discuss  how a single-photon wave packet with the stable size can be obtained.

According to the dispersion relation of the wave vector and angular frequency in Eq.~(\ref{ES}), the group velocity of photons in the $2$D photonic resonator array can be given as
 \begin{eqnarray}
\vec{v}=\nabla \omega(\vec{k})=(2J_x\sin(k_x),2J_y\sin(k_y)).
\label{group velocity}
\end{eqnarray}
Thus, if a single-photon wave packet propagates in the $2$D photonic resonator array with the group velocity $\vec{v}\sim(2J_x,0)\equiv (v_{x}=2J_x,v_{y}=0)$,  then such a single-photon wave packet is composed of modes $k_x\sim\pi/2$ and $k_y=0$. That is, the wave vector of this wave packet is $\vec{k}\sim(\pi/2,0)$. It is clear that there is no dispersion along $y$ direction when this wave packet propagates with the group velocity $\vec{v}\sim(2J_x,0)$, but this wave packet has broadening along the $x$ direction with its propagation.

To better understand the dispersion effect on the propagation of the wave packet in $2$D photonic resonator array,  we now assume that a single-photon wave packet is initially prepared to a state, for example,
\begin{equation}\label{eq:17}
|\psi_0\rangle=\sum_{m=-M}^{M}\alpha a_{n_c,m}^{\dag}|0,g\rangle,
\end{equation}
where all sites of photonic excitations in the wave packet form a straight line with $n=n_{c}<0$ parallel to the $y$-axis, the size $L_{y}$ of the wave packet along $y$-axis is assumed  to be $2M+1$, and the single-photon excitation at each lattice site $(n_{c}, m) $ with $m=-M,\cdots, M$ has the same probability $|\alpha|^2$ with the normalization $(2M+1)|\alpha|^2=1$. That is, the size of the single-photon wave packet is initially a straight line with the length $2M+1$ along $y$ axis and locates in the left of the origin in the $x$ axis. The state $|0,g\rangle$ denotes that the giant atom is the ground state $|g\rangle$ and the $2$D photonic resonator array is in its vacuum state $|0\rangle$.

Let us assume that the single-photon wave packet in Eq.~(\ref{eq:17}) freely propagates from the left side of the origin to the right side of the origin with the the group velocity  $\vec{v}\sim(2J_x,0)$. That is, we assume that the excited sites in $2$D photonic resonator array are far away from the giant atom or the  $2$D photonic resonator array and the giant atom are decoupled from each other during the photon propagation, the giant atom has a negligibly small or no effect on the wave packet in the $2$D photonic resonator array. Thus, the time evolution of both $2$D photonic resonator array and the giant atom is governed by their free Hamiltonian in Eq.~(\ref{H0}). Hereafter,  the time evolution of the  $2$D photonic resonator array is considered to be equivalent to the photon propagation.

\begin{figure}[tbp]
\centering
\includegraphics[width=8cm]{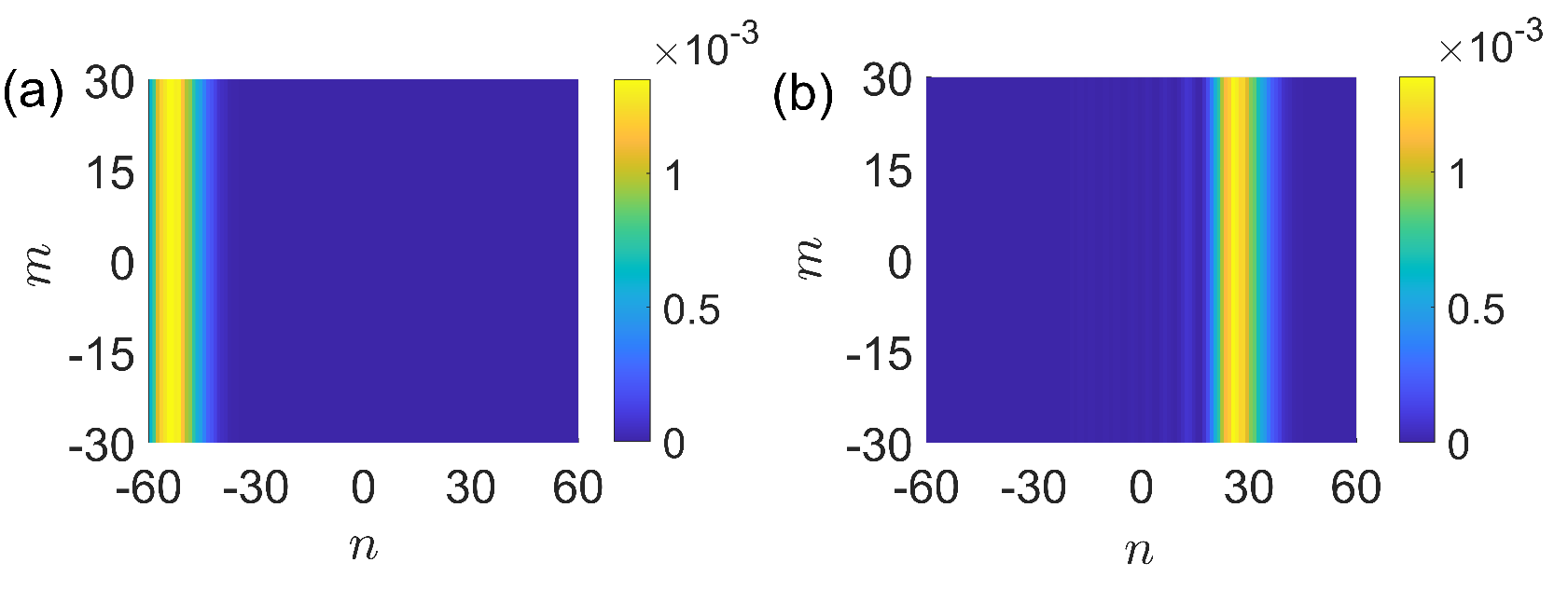}\nonumber\\
\includegraphics[width=8cm]{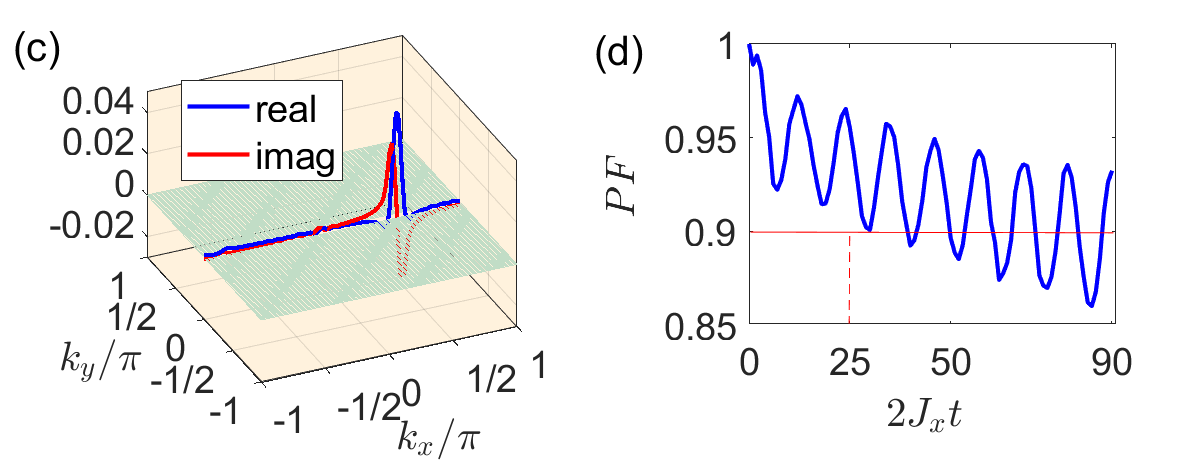}
\caption{(a) The photon distribution  $|\langle \psi_l |a^{\dag}_{n,m}|0,g\rangle|^2$ at each lattice site $(n,m)$ for  the wave packet state in Eq.~(\ref{psi0}). (b) The photon  distribution  $|\langle \psi_f(t)|a^{\dag}_{n,m}|0,g\rangle|^2$ at each lattice site $(n,m)$ for the wave packet $|\psi_{f}(t)\rangle$ with the evolution time $t=80/(2J_x)$. (c) The real part ${\rm Re}(\langle\psi_l|a^{\dag}_{k_x,k_y}|0,g\rangle)$ and imaginary part ${\rm Im}(\langle\psi_l|a^{\dag}_{k_x,k_y}|0,g\rangle)$ in the moment space  for the  wave packet in Eq.~(\ref{psi0}). (d) The propagating fidelity PF of the initial wave packet versus the time $t$. The parameters are set $J_y/(2J_x)=0.2$, $N=121$, $M=61$, $L_x=25$.  $v_{x}t=2J_x t=80$ for (b).}
\label{cavityT}
\end{figure}

It is clear that the wave packet has no broadening along $y$ direction due to $v_{y}=0$ when it propagates with the  group velocity $\vec{v}\sim(2J_x,0)$.  That is, the size $2M+1$ of the $y$ direction for this wave packet is not changed with the time evolution. The size of the $x$ direction is increased with the time evolution. To obtain the wave packet with a stable size, we now assume that the wave packet dynamically evolve with a time $T$ from the initial state in Eq.~(\ref{eq:17})  under the Hamiltonian $H_0$ given in Eq.~(\ref{H0}), then the initial wave packet evolves to
\begin{eqnarray}\label{eq:19-1}
|\psi_0^{\prime}\rangle&=& e^{-iH_0T}\sum_{m=-M}^{M}\alpha a_{n_c,m}^{\dag}|0,g\rangle, \nonumber\\
&=&\sum_{m=-M}^{M}\sum_{k_x,k_y}\frac{\alpha a_{k_x,k_y}^{\dag}e^{i\omega(k) T-k_x n_c-k_y m}}{\sqrt{(2M+1)(2N+1)}}|0,g\rangle\nonumber\\
&=&\sum_{m=-M}^M\sum_{n=-N}^{N}\alpha_n a_{n+n_c,m}^{\dag} |0,g\rangle
\end{eqnarray}
where $\alpha_n=\alpha e^{i(v_x-\omega_l)T}i^{n}\mathcal{J}_{n}(v_xT)$ and $\mathcal{J}_{n}(v_xT)$ denotes the Bessel function of the first kind of order $n$ with the variable $v_xT$.
Let us  truncate a part in the position space from the wave packet $|\psi_0^{\prime}\rangle$ in Eq.~(\ref{eq:19-1}) with the size $L_x\times L_y$ as schematically shown in Fig.~\ref{2DEnergyspectrum}(a),  the center line $n=n_{i}$ along $y$ direction of the initial wave packet in Eq.~(\ref{eq:17}) is now changed to  $n_a=n_c+v_xT$ for the truncated wave packet,  and the lattice site coordinate in $y$-axis of the truncated wave packet is not changed due to $v_{y}=0$ and still takes values $m=-M,\cdots, M$. The size $L_{x}$ is determined by the group velocity and is discussed below. We now name the truncated wave packet state  from Eq.~(\ref{eq:19-1})  as $|\psi_1\rangle$ and denote its normalized form as
\begin{equation}\label{eq:18}
|\psi_1\rangle=\sum_{m=-M}^{M}\sum_{n=-(L_{x}/2)+n_{a}}^{(L_x/2)+n_{a}}
\alpha_{n}a_{n,m}^{\dag}|0,g\rangle.
\end{equation}
We use Eq.~(\ref{eq:18}) as a new initial wave packet state and let the wavefunction  evolve the time $T$ for the second iteration to obtain the new wavefunction
\begin{equation}\label{eq:21-1}
|\psi^{\prime}_1\rangle=\exp(-iH_{0}T)|\psi_1\rangle,
\end{equation}
which can be derived with the same method as for Eq.~(\ref{eq:19-1}).
%Based on the Schr\"{o}dinger equation $i\frac{\partial}{\partial t} |\psi_1(t)\rangle = H_0' |\psi_1(t)\rangle$, and $\alpha(t)$ satisfy
% \begin{eqnarray}\label{eq:19}
%i\dot{\alpha}_{n,m}&=&\omega_l\alpha_{n,m}
%-J_x(\alpha_{n+1,m}+\alpha_{n-1,m})\nonumber \\
%&&-J_y(\alpha_{n,m+1}+\alpha_{n,m-1}).
%\end{eqnarray}}
Then we truncate another wave packet $|\psi_2\rangle$ from the wavefunction in Eq.~(\ref{eq:21-1}) with the same size $L_x\times L_y$ as for the wavefunction in Eq.~(\ref{eq:18}), but the center line is now changed to  $n_b=n_c+2v_xT$. We iterate such procedure many  times, e.g., $l$ times. That is, the evolution time for each iteration is assumed to be $T$, the size of the truncated wave packet is $L_x\times L_y$,  but the center line along $y$ direction for the truncated wave packet should add  $v_xT$. The iteration is finished until the truncated wave packet with the size $L_x\times L_y$ has a stable fidelity discussed below in Eq.~(\ref{PF}). We can assume that the formal solution of the truncated  wave packet $|\psi_L\rangle$ with $l$ iterations and the size $L_x\times L_y$ is written as
\begin{eqnarray}
|\psi_l \rangle=\sum_{m=-M}^{M}\sum_{n=-(L_{x}/2)+n_{f}}^{(L_x/2)+n_{f}}\beta_{n,m}e^{ik_{c,x}n+ik_{c,y}m}a_{n,m}^{\dag}|0,g\rangle, \nonumber\\
\label{psi0}
\end{eqnarray}
where the parameters $\beta_{n,m}$  satisfy the normalization condition $\sum_{m,n}|\beta_{n,m}|^2=1$, and are determined by $\alpha$ and the total time $T_{\rm total}=l T$  of $l$ iterations. The center line $n_{f}$ along $y$ direction for finally truncated wave packet should be $n_{f}=n_{c}+v_{x}T_{\rm total}$. In our calculation, we assume $n_{c}=-1048$ and $T=25/(2J_x)$.  We find that the size of $L_{x}$ is about $25$ with the given parameters when the wave packet reaches nearly stable with $n_{f}=-48$. The photon distribution $|\beta_{n,m}|^2$ of the wave packet in the position space is shown in Fig.~\ref{cavityT}(a).
In the following discussion, for the convenience of analyzing the scattering process, the values of N and M are set much larger than the propagating distance of the wave packet.
 In Fig.~\ref{cavityT}(c), we plot the real and imaginary parts of the initial wave packet $|\psi_l\rangle$ in the momentum space. It confirms that photon distribution is confined to be around the central wave vector $\vec{k}_{c}$ with $k_{c,x}\sim\pi/2$ and $k_{c,y}=0$.
We further study the time evolution of the wave packet with an evolution time $t$ when the initial state is expressed as $|\psi_l\rangle$ in Eq.~(\ref{psi0}), then we find that the final state
\begin{equation}\label{BG}
|\psi_f(t)\rangle=\exp(-iH_0t)|\psi_l\rangle
\end{equation}
 with an evolution time $t=80/(2J_x)$ exhibits localization in the position space and can propagate to the designated position with a group velocity  $\vec{v}\sim(2J_x,0)$ as show in Fig.~\ref{cavityT}(b). Its main photon distribution is concentrated in $m\in\{-M,\cdots, M\}$ and $n\in\{v_{x}t+n_{f}-(L_{x}/2),\cdots,v_{x}t+n_{f}+(L_{x}/2)\}$. Hereafter, we call the state in Eq.~(\ref{BG})  as  ``background'' state of the photons. This state describes the free evolution of the single-photon wave packet state in the 2D resonator array  without the coupling to the atom.

To describe the distortion degree of the wave packet during the propagation, we here define the propagating fidelity (${\rm PF}$)~\cite{MichaelMurphy}, which is inner product of the translational wave packet $D(t) |\psi_l\rangle$ and the evolved wave packet $|\psi_f(t)\rangle$ with an evolution time $t$
 \begin{eqnarray}
{\rm PF}=|\langle \psi_f(t)|D(t)|\psi_l\rangle|^2,
\label{PF}
\end{eqnarray}
where
 \begin{eqnarray}
D=\sum_{m=-M}^M\sum_{n=-(L_{x}/2)+n_{f}}^{(L_{x}/2)+n_{f}}a^{\dag}_{n+v_xt,m}a_{n,m}
\label{D1}
\end{eqnarray}
is the translational operation on the initial wave packet with a translational distance $v_xt=2J_{x}t$.  When ${\rm PF}=1$, we say that the wave packet perfectly propagates, otherwise when ${\rm PF}=0$, we say that the wave packet is completely distorted. In Fig.~\ref{cavityT}(d), we  have exhibited the ${\rm PF}$ of the wave packet versus the time $t$. It shows that the ${\rm PF}$ decreases with the time evolution and oscillates periodically. If the ${\rm PF}$ of the initial wave packet remains above 0.9 after propagating a distance $L_x$ as shown by the red line in Fig.~\ref{cavityT}(d),
then we say that the wave packet has a stable size, which can also serve as an incident light source for studying light scattering.
 Below, we will study the scattering of the single-photon wave packet, which is initially prepared to the state given in Eq.~\eqref{psi0} and has  given group velocity, e.g., $\vec{v}\sim(2J_x,0)$ or $\vec{v}\sim(2J_x,2J_y)$.

\section{Scattering of Single-photon Wave Packet by a small atom}
\label{Propagation and Scattering of Wave Packet}

Let us now study the scattering of the single-photon wave packet by a small atom, which is consider as a point particle. That is, we assume that the small atom is coupled to the photonic lattice at the point $(0,0)$ and $\omega_a=\omega_l$. Thus, the coupling strength in Eq.~\eqref{Gk} between the small atom and $2$D photonic resonator array in the momentum space is simplified to $G(\vec{k})=g$.

\begin{figure}[tbp]
\centering
\includegraphics[width=8cm]{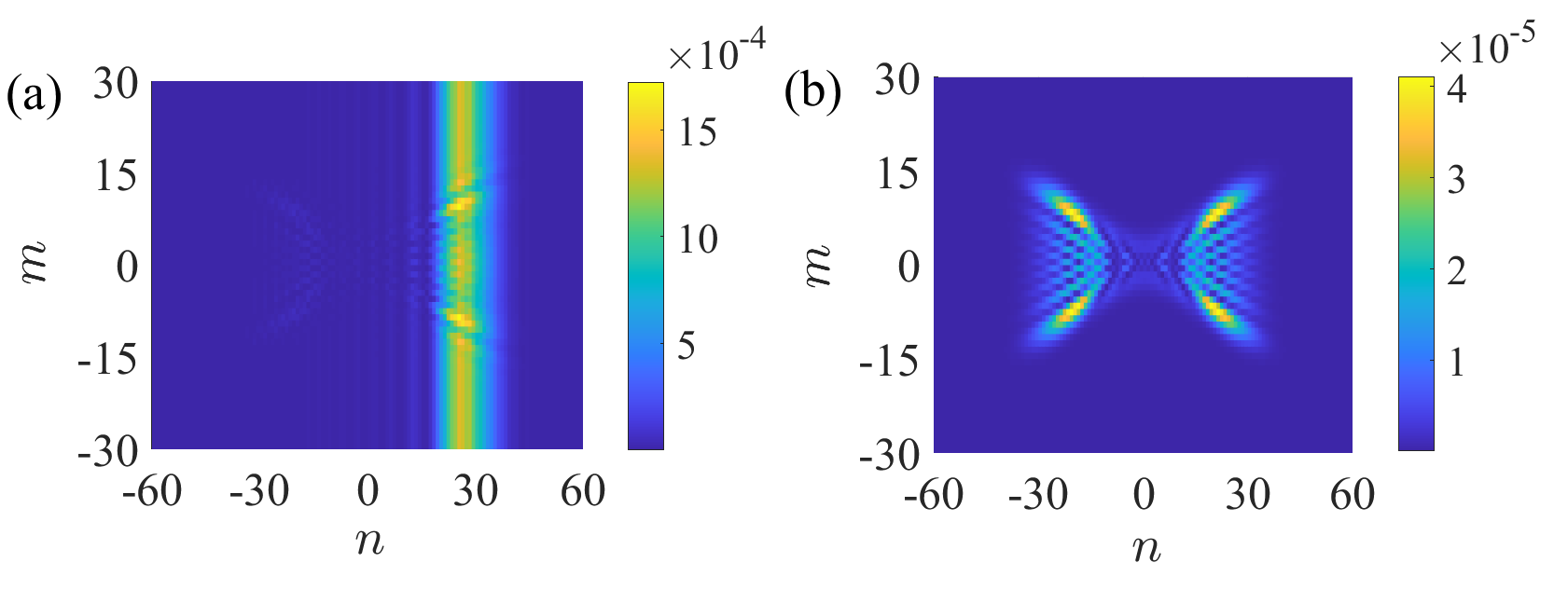}\nonumber\\
\includegraphics[width=8cm]{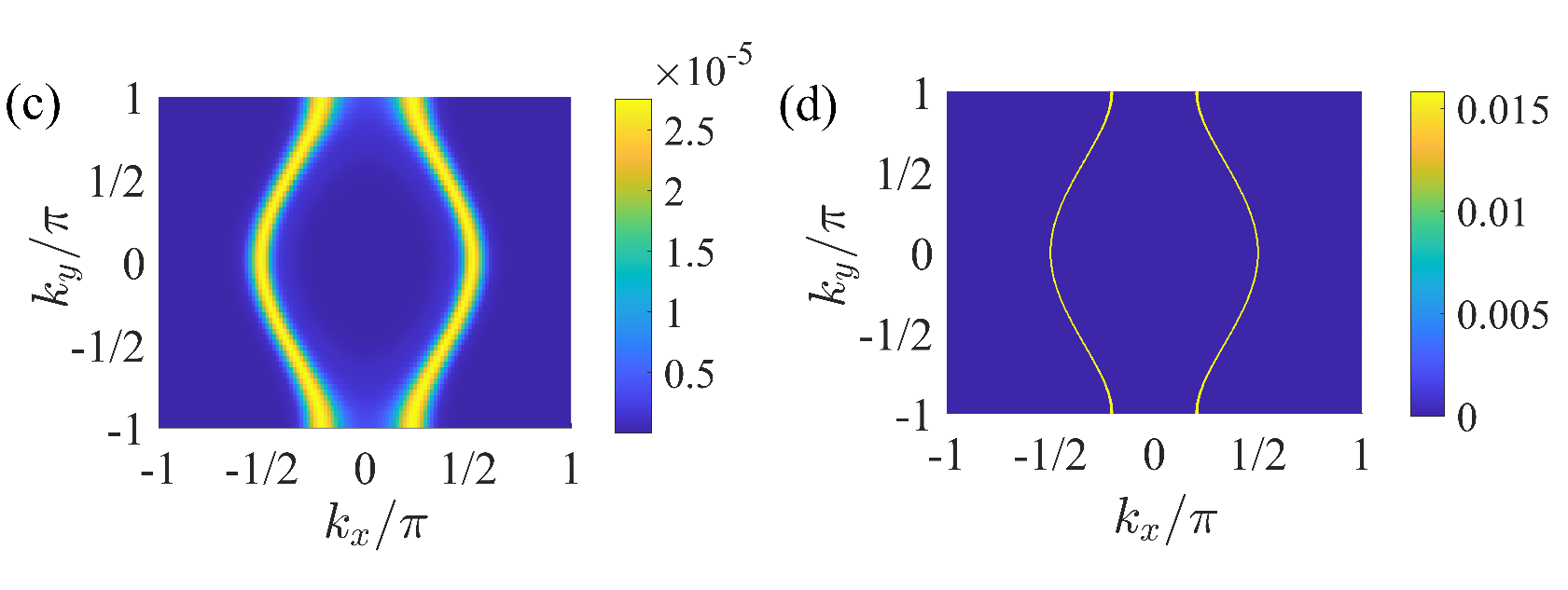}
\caption{(a) and (b) The photon distributions of the scattered wave packets with $|\langle \psi_s(t)|a^{\dag}_{n,m}|0,g\rangle|^2$ and without $|(\langle\psi_s(t)|-\langle\psi_f(t)|)a^{\dag}_{n,m}|0,g\rangle|^2$ the ``background" for the small atom, respectively.
(c) The photon distribution  $|(\langle\psi_s(t)|-\langle\psi_f(t)|)a^{\dag}_{k_x,k_y}|0,g\rangle|^2$ of the scattering wave packet without the ``background" in the momentum space for the small atom. (d) The distribution of the photon scattering expression $|\langle g,\vec{k}_f|S-1|\vec{k}_i,g\rangle|^2$ for the small atom.
The main parameters are consistent with Fig.~\ref{cavityT}.
The other parameters are set as: $n_0=m_0=0$, $g=2J_x$, $\omega_l=\omega_a=0$.  $2J_x t=80$ for (a), (b) and (c).}
\label{SMatom}
\end{figure}

To study the scattering properties of the system, we assume that the atom is initially in  the ground state $|g\rangle$ and the photonic resonator array is initially prepared to the  wave packet state $|\psi_l\rangle$ given in Eq.~\eqref{psi0} and shown in Fig.~\ref{cavityT}(a). The single-photon wave packet propagates in the group velocity $\vec{v}\sim(2J_x,0)$. With an evolution time $t$,  the system is changed from the initial state $|\psi_l\rangle$ to final state $|\psi_s(t)\rangle=\exp(-iHt)|\psi_l\rangle$, where $H$ is the total Hamiltonian of the small atom and the photonic resonator array given in Eq.~\eqref{H}. As shown in Fig.~\ref{SMatom}(a),  the photonic distribution probabilities $|\langle \psi_s(t)|a^{\dag}_{n,m}|0,g\rangle|^2$ for the wave packet state $|\psi_s(t)\rangle$ is plotted.  It clearly shows  that the atom has negligibly small effect on the part of the incident wave packet far from the atom during propagation, but the part of the incident wave packet close to the atom undergoes significant scattering. To clearly show the influence of the scatterer to the wave packet, we subtract the ``background''  as shown in Fig.~\ref{cavityT}(b) from Fig.~\ref{SMatom}(a), and calculate the photon distribution probabilities  $|(\langle\psi_s(t)|-\langle\psi_f(t)|)a^{\dag}_{n,m}|0,g\rangle|^2$ as shown in Fig.~\ref{SMatom}(b) with the state $|\psi_f(t)\rangle=\exp(-iH_{0}t)|\psi_l\rangle$.  It shows that the scattered photon distributes symmetrically around the small atom. The symmetry originates from the energy spectrum of the system as shown in Fig.~\ref{2DEnergyspectrum}(b). As shown in Fig.~\ref{SMatom}(c), we plot the photon distribution probabilities $|(\langle\psi_s(t)|-\langle\psi_f(t)|)a^{\dag}_{k_x,k_y}|0,g\rangle|^2$ in the momentum space. It shows that the photons distribute around constant-energy of the 2D resonator array.  In practice, the method of ''background" elimination can be attributed to the influence of the second term in the $S$-matrix in Eq.~\eqref{Tmatrix}. Therefore, in Fig.~\ref{SMatom}(d), we plot the photon distribution probabilities $|\langle g,\vec{k}_f|S-1|\vec{k}_i,g\rangle|^2$ in momentum space versus $k_{f,x}$ and $k_{f,y}$ by using the analytic result in Eq.~\eqref{Tmatrix} with $k_{i,x}=\pi/2$ and $k_{i,y}=0$. The agreement of numerical and analytical results shows the rationality of our approach for dynamically simulation for the scattering process.

We note that  the scattering probability of the photons in one-dimensional systems can be significantly modulated by adjusting the frequency of the atoms and the coupling strength between the atoms and the waveguide~\cite{Zhou,Nie}. However, in 2D systems, the photons scattered by a small atom uniformly  distribute around the constant-energy surface of the system. Therefore, we find that the distribution modulation of scattered photons by the small atom coupled to $2$D resonator array has the significant limitation.
In Appendix.~\ref{Appendixb}, we study the distribution of scattered photons by the giant atom coupled to in the 2D photonic resonator array via a few discrete connection points when the system evolves a time $t$ from the initial state $|g\rangle\otimes|\psi_l\rangle$. We find that the scattered photons no longer uniformly distribute on the constant-energy surface due to photon interference effects between the coupling points, and the distribution probabilities of photons present an interference pattern in the constant-energy surface.
Therefore, in the following, we develop the giant atom setup and design on demand scattering process by optimizing  the coupling between the giant atom and the 2D photonic resonator array.

\section{Single-Photon Scattering by a Giant Atom coupled to 2D resonator array via 1D spatial points}
\label{Controllable}

\subsection{Optimization Function}

We now turn to study the photon scattering by a giant atom in 2D resonator array. The nonlocal coupling between the giant atom and 2D resonator array via multiple spatial points results in  nontrivial phase accumulations of the propagating field, which induce the photon interference between different coupling points. This interference provides a possible way to optimally design the photon scattering, which is different from that in the system of many atoms~\cite{POGuimond,DZXu}. Here, we first consider that all of the coupling points aline along the $y$-axis from the lattice site $(0, -M)$ to the site $(0, M)$, i.e., the giant atom-resonator array coupling points are symmetric with $y=0$. Then Eq.~\eqref{Gk} can be simplified to
 \begin{eqnarray}
G(\vec{k})=\frac{2g}{\Lambda}\left\{\sum_{m_1=0}^{M_1}\lambda_{m_1}\cos(m_1 k_y)\right\}.
\label{Gk2}
\end{eqnarray}
Here, we set $\lambda_{0,0}\rightarrow 2\lambda_0$  and $\lambda_{0,m_{1}}\equiv \lambda_{m_1}$ ($m_1=1,2,...,M_1$). We assume $\lambda_{m_1}=\lambda_{-m_{1}}$ when Eq.~(\ref{Gk2}) is derived from  Eq.~\eqref{Gk}.
Thus, the  parameter $2g\lambda_{0}/\Lambda$  is the coupling strength  between the giant atom and the resonator array at the site $(0,0)$, and the parameter $g\lambda_{m_1}/\Lambda$ corresponds to the coupling strength between the giant atom and waveguide at the site $(0,\pm m_1)$.

\begin{center}
\begin{table*}[htbp]
\centering
\caption{The values of $\xi_{m_1,m_2}$ with parameters $J_y/(2J_x)=0.2$. }
\begin{tabular}{|c|c|c|c|c|c|c|c|c|}
\hline
$\xi_{m_1,m_2}$ & $m_2=0$ & $m_2=1$ & $m_2=2$ & $m_2=3$ & $m_2=4$ & $m_2=5$ & $m_2=6$ & $m_2=7$ \\
\hline
  $m_1=0$ & -1.9985i & 0.1470i  & -0.0598i & 0.0173i & -0.0061i & 0.00207i & -0.00073i & 0.000259\\
  \hline
  $m_1=1$ & \  & -0.6298i & 0.08215i & -0.03295i & 0.009679i & -0.0034i & 0.001163i & -0.0004117i  \\
  \hline
  $m_1=2$ & \  & \  & -0.5208i & 0.07454i & -0.0303i & 0.0088i & -0.00309i & 0.001082i  \\
  \hline
  $m_1=3$ & \  & \  & \  & -0.6003i & 0.0736i & -0.02996i & 0.008663i & -0.003048i \\
  \hline
  $m_1=4$ & \  & \  & \  & \  & -0.59997i & 0.0735i & -0.02992i & 0.008648i\\
  \hline
  $m_1=5$ & \  & \  & \  & \  & \  & -0.5994i & 0.07351i & -0.02991i \\
  \hline
  $m_1=6$ & \  & \  & \  & \  & \  & \  & -0.59993i & 0.07351i\\
  \hline
  $m_1=7$ & \  & \  & \  & \  & \  & \ & \  & -0.59993i \\
\hline
\end{tabular}
  \label{table1}
\end{table*}
\end{center}

We know that the single-photon wave packet is scattered by the giant atom to all directions with different momentums in the momentum space.
However,  if we can  manipulate the coupling strengths between the giant atom and resonator array via the coupling points, then we can engineer the scattering photons to the expected state.  In the following,  we show how we can obtain an expected target scattering state, which is composed of several sets of momentum modes by well engineering the coupling constants of different coupling points. Hereafter, we refer to these momentums as expected momentums.
In the preceding analysis, we focused on a symmetric coupling configuration, which results in the symmetry for expected momentums.
Therefore, each set, e.g.,  the $j$th set of momentum modes in the target state,  consists of  four different momentum modes  $(k_{{\rm xtar,j}}, k_{{\rm ytar,j}}),~(-k_{{\rm xtar,j}}, k_{{\rm ytar,j}}),~( k_{{\rm xtar,j}}, -k_{{\rm ytar,j}}),$ $(-k_{{\rm xtar,j}},-k_{{\rm ytar,j}})$, which are symmetrical distribution in four quadrants of the $xy$-plane.   The integer number $j$ labels the set of modes, $k_{{\rm xtar,j}}$ and $-k_{{\rm xtar,j}}$ ($k_{{\rm ytar,j}}$ and $-k_{{\rm ytar,j}}$) are assumed to be two centers for the $j$th set of modes in x (y) direction of the momentum space. They satisfy the relation $E_i-E(\vec{k}_{{\rm tar,j}})=0$ with the energy $E_i$ of a single photon in the initial state as shown in Eq.~(\ref{Smatrix}).

To achieve the target scattering state with $j$ sets of momentum modes, we define an optimization function:
 \begin{eqnarray}
Q\equiv Q(k_{{\rm ytar,j}},k_{j},\lambda_{0},\lambda_{1}...\lambda_{M_1})=\frac{\Pi_{j}P_{j}}{O-\Sigma_{j}P_{j}},
\label{optimization1}
\end{eqnarray}
with
\begin{eqnarray}
P_{j}(k_{{\rm ytar,j}},k_{j},\{\lambda_{m_1}\}) =\int_{-\pi}^{\pi}dk_{{\rm xf}}\int_{\Delta_{-}}^{\Delta_+}dk_{{\rm yf}}\left|S_{f,i}\right|^2
\label{optimization2}
\end{eqnarray}
and
\begin{eqnarray}
O(k_{{\rm ytar,j}},k_{j},\{\lambda_{m_1}\})=\int_{-\pi}^{\pi}dk_{{\rm xf}}
\int_{0}^{\pi}dk_{{\rm yf}}\left|S_{f,i}\right|^2.
\label{optimization3}
\end{eqnarray}
The function  $P_{j}\equiv P_{j}(k_{{\rm ytar,j}},k_{j},\lambda_{0},...,\lambda_{M_1})\equiv P_{j}(k_{{\rm ytar,j}},k_{j},\{\lambda_{m_1}\}) $ denotes the probability that the photon momentums of the scattering state are in the range $k_{x}\in (-\pi, \pi)$ and  $k_{y}\in (\Delta_-, \Delta_+)$ with $\Delta_{\pm}=k_{{\rm ytar,j}}\pm k_j/2$ and  $S_{f,i}=\langle g, \vec{k}_f|S-1|\vec{k}_i,g\rangle$. Here, $k_j$ is the width of the $j$th set of the modes in $y$ direction.  However, the function $O\equiv O(k_{{\rm ytar,j}},k_{j},\lambda_{0},...,\lambda_{M_1})\equiv O(k_{{\rm ytar,j}},k_{j},\{\lambda_{m_1}\})$ denotes the probability that the momentums of the scattering photons are in the range $k_{x}\in(-\pi, \pi)$ and $k_{y}\in (0, \pi)$. We note that we only consider the case of $0<k_{{\rm ytar,j}}<\pi$ due to the symmetric distribution of the target scattering state  with the reflectional symmetric axis $y=0$. Then we have $k_{{\rm ytar,j}}-k_j/2\geq0$ and $k_{{\rm ytar,j}}+k_j/2\leq\pi$.

In the optimization function $Q$,  the continuous multiplication of $\{P_j\}$ guarantees that photons in the target scattering state have the expected momentum. The function $O-\Sigma_{j}P_j$ is the probability that photons in the target scattering state have no the expected momentum.  To obtain the optimization function $Q$, we need to calculate the self-energy function $\Sigma$ given in Eq.~\eqref{self-energy} and  the coupling strength in Eq.~\eqref{Gk2} which can be expanded  as
\begin{eqnarray}
&&|G(\vec{k})|^2=\frac{4|g|^2}{\Lambda^2}\left[\sum_{m_1=0}^{M_1}|\lambda_{m_1}|^2\cos^2(m_1k_y) \right.\\
&&+\left. \sum_{m_1=0}^{M_1-1}\sum_{m_2=m_1+1}^{M_1}2 {\rm Re}(\lambda^*_{m_1}\lambda_{m_2})\cos(m_1k_y)\cos(m_2k_y)\right].\nonumber
\label{Sigma12}
\end{eqnarray}
with the normalized constant $\Lambda=\left(\sum_{m_1=0}^{m_{1}=M_{1}}2|\lambda_{m_1}|\right)^2$. If we define
\begin{eqnarray}
\xi_{m_1,m_2}=
\frac{2J_x}{4\pi^2}\int dk_xdk_y\frac{\cos(m_1k_y)\cos(m_2k_y)}{\omega_l-2J_y-\omega(\vec{k})+i0^+},
\label{IntSigma12}
\end{eqnarray}
then the self-energy function $\Sigma$  in Eq.~\eqref{self-energy} can be simplified to
\begin{eqnarray}
\Sigma &=&\frac{4|g|^2}{2J_x \left(\sum_{m_1}|\lambda_{m_1}|\right)^2}
\left\{\sum_{m_1=0}^{M_1}|\lambda_{m_1}|^2\xi_{m_1,m_1}  \right.\nonumber \\
 &+& \left.\sum_{m_1=0}^{M_1-1}\sum_{m_2=m_1+1}^{M_1}2{\rm Re}(\lambda^*_{m_1}\lambda_{m_2})\xi_{m_1,m_2} \right\}.
\label{Sigmax}
\end{eqnarray}
 That is, once $\xi_{m_1,m_2}$ and $\{\lambda_{m_1}\}$ are given, then we can obtain $\Sigma$. For example, in Table~\ref{table1}, a set of $\xi_{m_1,m_2}$ is given for $15$ coupling points by numerical calculation when  $J_y/(2J_x)=0.2$. By optimizing coupling strengths $\{\lambda_{m_1}\}$, we can obtain the optimized function $Q$ and target scattering state.

To obtain the target scattering state with expected photon distributions, the optimization function $Q$ should be as large as possible by optimizing coupling strengths $\{\lambda_{m_1}\}$ for any number $2M_1+1$ of the coupling points.  Such maximization of the  optimization function $Q$ can be converted to a convex optimization problem. Thus, we have to introduce optimization algorithms. For a relatively simple optimization function with few scattering modes, the gradient descent~\cite{Ruder} is sufficient to deal with it. However, when the optimization function is composed of many modes, the particle swarm optimization~\cite{Kennedy} is an optional effective optimization algorithm for obtaining the optimization function $Q$.

\subsection{Controllable Symmetric scattering of  single-photon wave packet by a giant atom}
\label{Dynamic Scattering form a Giant Atom}

 \begin{figure}[tbp]
\centering
\includegraphics[width=8cm]{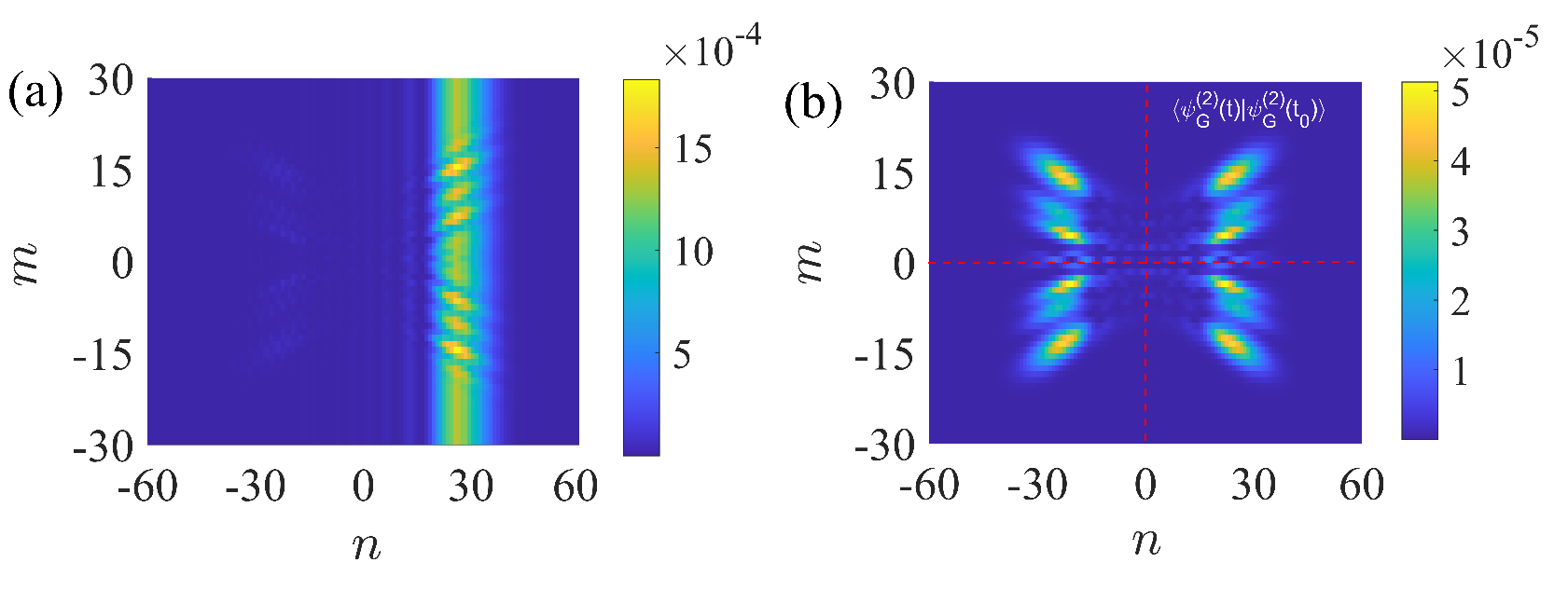}\nonumber\\
\includegraphics[width=8cm]{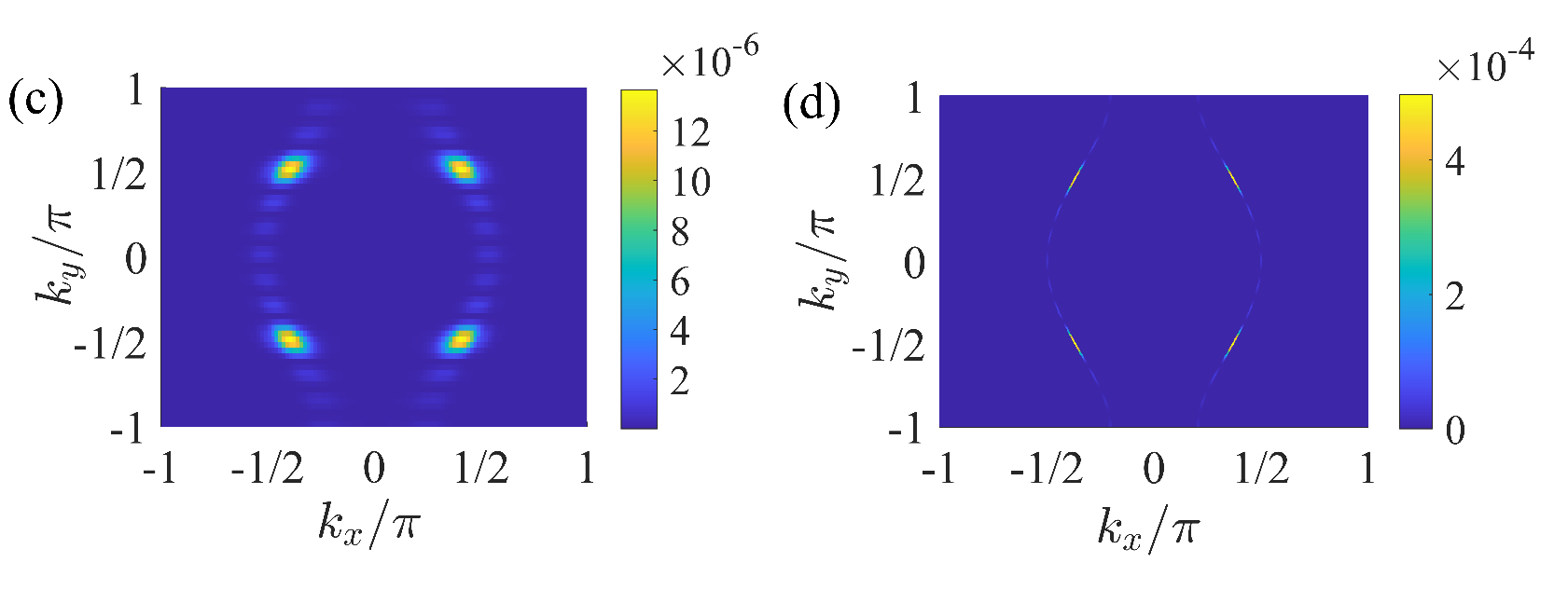}\nonumber\\
\includegraphics[width=8cm]{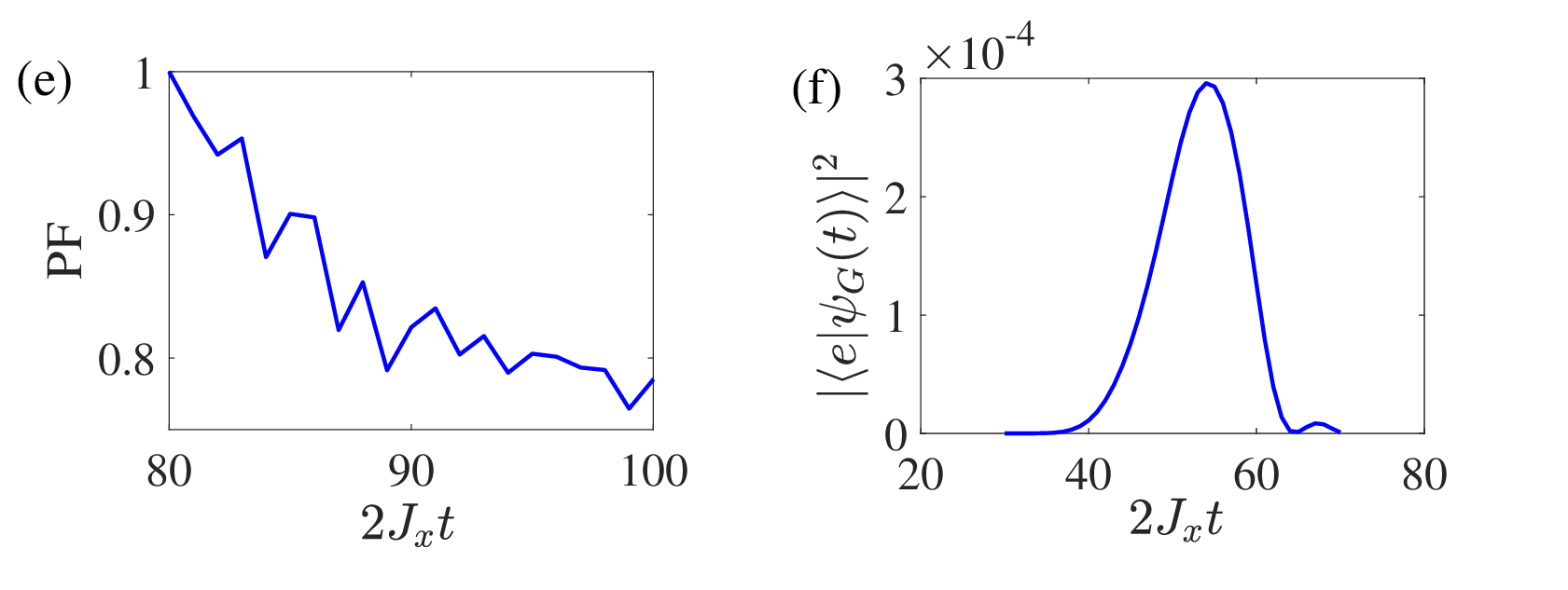}
\caption{(a) and (b) The photon distributions $|\langle\psi_G(t)|a^{\dag}_{n,m}|0,g\rangle|^2$ and $|(\langle\psi_G(t)|-\langle\psi_f(t)|)a^{\dag}_{n,m}|0,g\rangle|^2$ of the scattering wave packets with and without the ``background" for the giant atoms, respectively.
(c) The photon distribution $|(\langle\psi_G(t)|-\langle\psi_f(t)|)a^{\dag}_{k_x,k_y}|0,g\rangle|^2$ of the scattering wave packet without the background in momentum space for the giant atom.
(d) The distribution of the photon scattering expression $|\langle g,\vec{k}_f|S-1|\vec{k}_i,g\rangle|^2$ for the giant atom.
(e) The PF of the scattering wave packet versus the time $t$.
(f) The excited state population $|\langle e,0|\psi_G(t)\rangle|^2$ of the giant atom versus the time $t$.
The main parameters are consistent with Fig.~\ref{SMatom}.
The other parameters are set as: $M_1=7$, $k_{ytar,1}=\pi/2$, $k_1=\pi/7$.
 $2\lambda_{0}: \lambda_1: \lambda_2: \lambda_3: \lambda_4:  \lambda_5:\lambda_6:\lambda_7=1:-0.00088:-2.0547:-0.0476:2.1248:0.0026:-2.6811:-0.2156$. $2J_x t=80$ for (a), (b) and (c).
$t_0=80/(2J_x)$, $L'_x=20$ and $L'_y=18$ for (e).}
\label{giantGD2}
\end{figure}

 \begin{figure}[tbp]
\centering
\includegraphics[width=8cm]{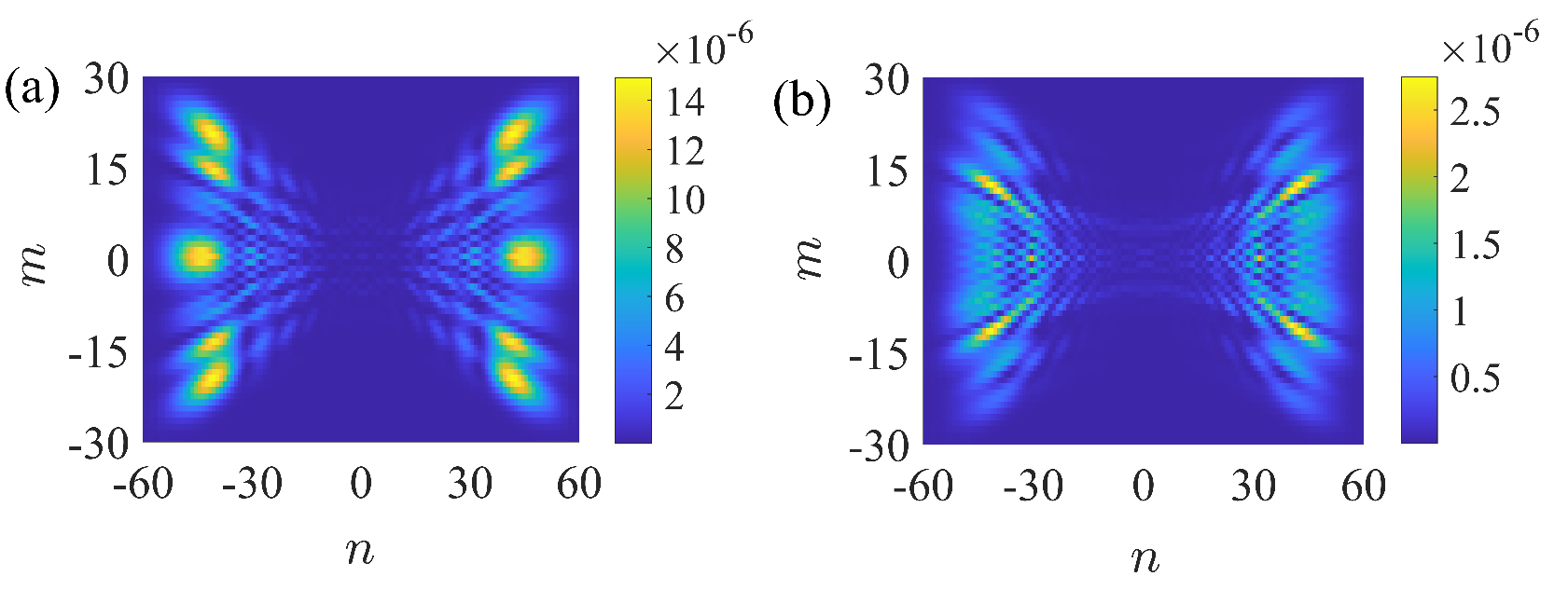}\nonumber\\
\includegraphics[width=8cm]{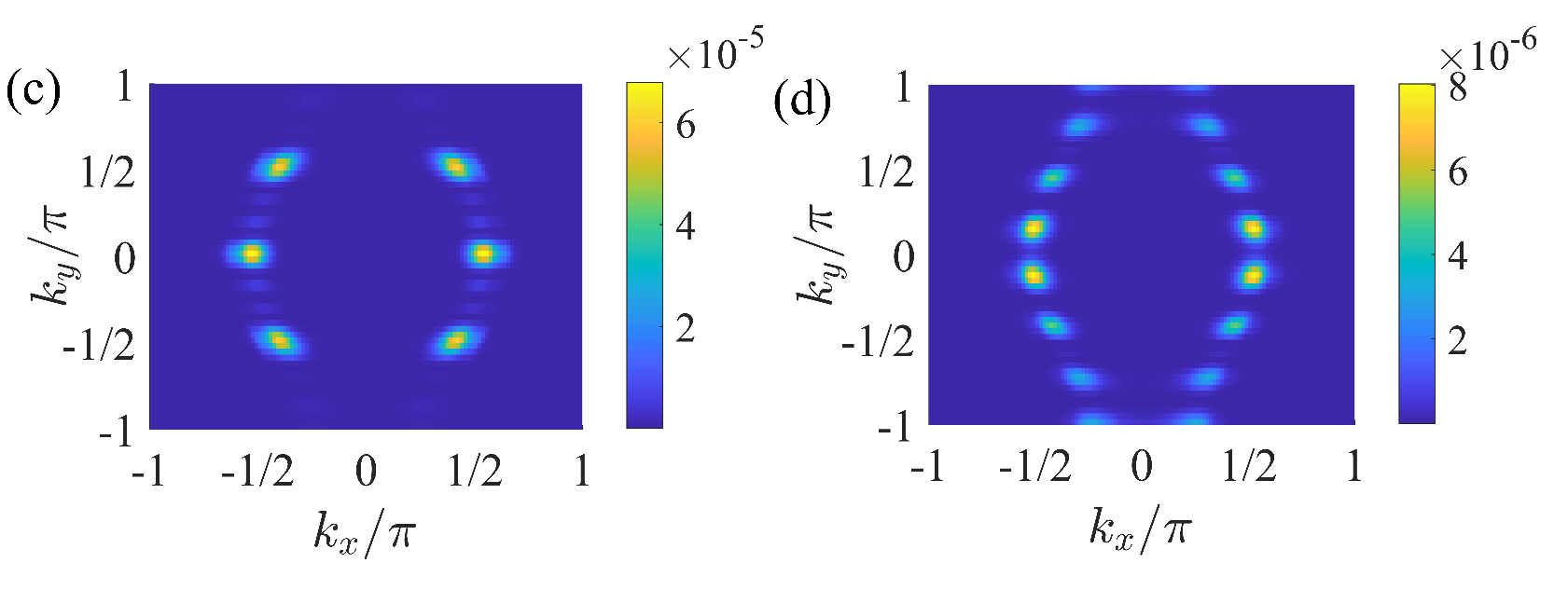}\nonumber\\
\includegraphics[width=8cm]{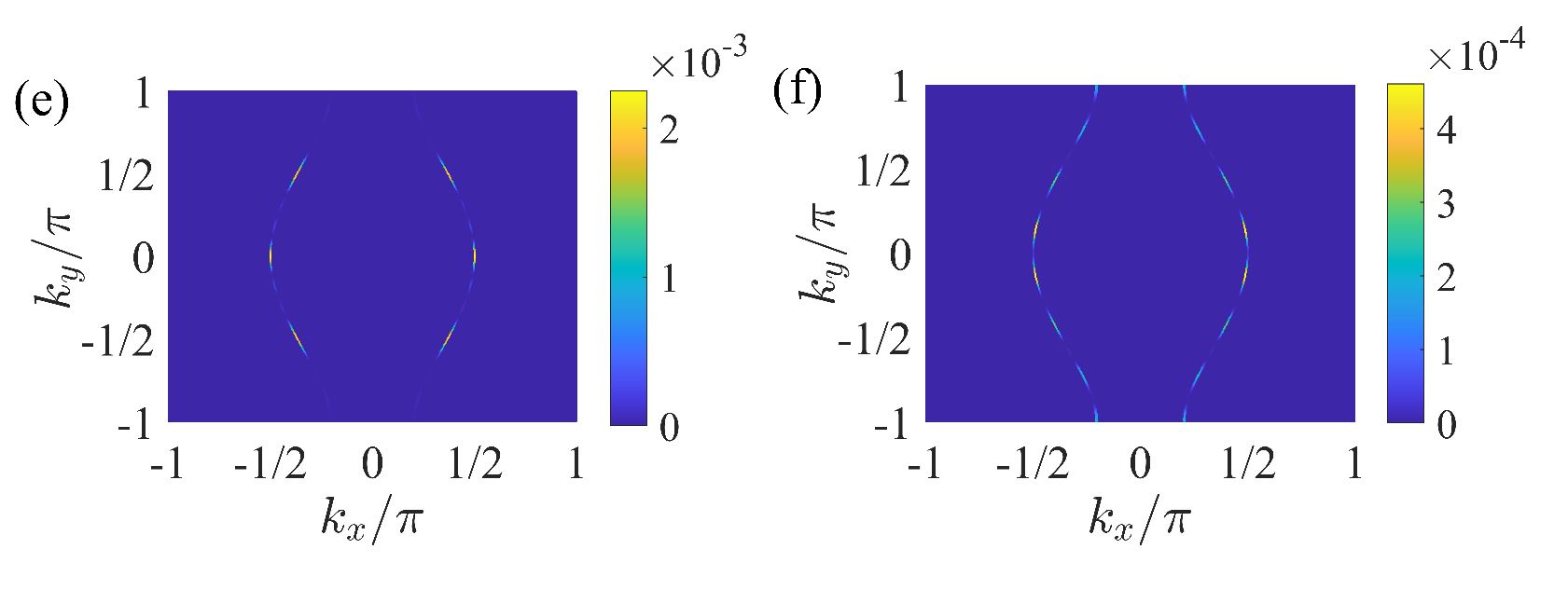}
\caption{(a) and (b) The photon distributions $|(\langle\psi_G(t)|-\langle\psi_f(t)|)a^{\dag}_{n,m}|0,g\rangle|^2$ of the scattering wave packets without the ``background" for two different types of giant atoms. (c) and (d) The photon distributions $|(\langle\psi_G(t)|-\langle\psi_f(t)|)a^{\dag}_{k_x,k_y}|0,g\rangle|^2$ of the scattering wave packets without the ``background" in momentum space for two different types of giant atoms.
(e) and (f) The distributions of the photon scattering expression $|\langle g,\vec{k}_f|S-1|\vec{k}_i,g\rangle|^2$ from two different types of giant atoms.
The main parameters are consistent with Fig.~\ref{SMatom}.
The other parameters are set as: $M_1=7$, $2J_x t=100$ for (a), (b), (c) and (d).
$k_{ytar,1}=\pi/28$, $k_{ytar,2}=\pi/2$, $k_1=\pi/14$, $k_2=pi/7$, $2\lambda_{0}: \lambda_1: \lambda_2: \lambda_3: \lambda_4:  \lambda_5:\lambda_6:\lambda_7=1:-1.4332:-3.2858:-1.3093:1.1984:-1.3141:-3.2387:-0.9584$ for (a), (c) and (e).
$k_{ytar,1}=\pi/7$, $k_{ytar,2}=3\pi/7$, $k_{ytar,3}=5\pi/7$, $k_{ytar,4}=27\pi/28$, $k_1=k_2=k_3=\pi/7$, $k_4=pi/14$, $2\lambda_{0}: \lambda_1: \lambda_2: \lambda_3: \lambda_4:  \lambda_5:\lambda_6:\lambda_7=1:0.3509:0.2903:-0.0347:0.00684:0.1582:-0.1743:-0.9628$ for (b), (d) and (f).}
\label{giantT24}
\end{figure}

By using the optimization function $Q$ and the gradient descent algorithm, we can obtain a set of optimal coupling parameters $\{\lambda_{m_1}\}$ to achieve the symmetric target scattering states. For instance, we assume that the giant atom is coupled to the $2$D photonic resonator array through $15$ coupling points, the system is initially in the state given in Eq.~\eqref{psi0}, and the photons propagate with the group velocity $\vec{v}=(2J_{x},0)$. If we expect that the target scattering state has a stable configuration in the momentum space as shown in Fig.~\ref{giantGD2}(c) after the single-photon wave packet is scattered by the giant atom with a time $t$, then we can  use the optimization function $Q$ and the gradient descent algorithm to obtain the  optimal  coupling parameters,  which satisfy the condition $2\lambda_{0}: \lambda_1: \lambda_2: \lambda_3: \lambda_4: \lambda_5:\lambda_6:\lambda_7=1:-0.00088:-2.0547:-0.0476:2.1248:0.0026:-2.6811:-0.2156$.

For above optimized coupling strengths, let us show the behavior of the symmetric target scattering state $|\psi_G(t)=\exp(-iHt)|\psi_l\rangle$ with the scattering time $t$ in the position space, where $|\psi_l\rangle$ is given in Eq.~\eqref{psi0} and the total Hamiltonian $H$ is given in Eq.~\eqref{H} for $15$ coupling points with above optimized coupling strengths. In the position space,  the photon distribution probabilities $|\langle\psi_G(t)|a^{\dag}_{n,m}|0,g\rangle|^2$ are plotted in Fig.~\ref{giantGD2}(a). In  Fig.~\ref{giantGD2}(b),  the photon distribution  probabilities $|(\langle\psi_G(t)|-\langle\psi_f(t)|)a^{\dag}_{n,m}|0,g\rangle|^2$ are plotted when we subtract the ``background"  state $|\psi_f(t)\rangle=\exp(-iH_0t)|\psi_l\rangle$ with $H_0$ given in Eq.~\eqref{H0}. Both Fig.~\ref{giantGD2}(a) and Fig.~\ref{giantGD2}(b) clearly show that the configuration of initial state $|\psi_l\rangle$ in Eq.~\eqref{psi0} shown in Fig.~\ref{cavityT}(a) is changed by the giant atom. We can also find that the target scattering single-photon wave packet in  Fig.~\ref{giantGD2}(c) is composed of the set of the modes $\{(\arccos(J_y/J_x),\pi/2) ,(-\arccos(J_y/J_x),\pi/2)$, $(\arccos(J_y/J_x),-\pi/2), (- \arccos(J_y/J_x),-\pi/2)\}$. The group velocities $v_x=2J_x$ and $v_y=0$ of the incident wave packet  are changed  to $v_x=\pm2\sqrt{1-(J_y/J_x)^2}J_x$ and  $v_y=\pm2J_y$ after the wave packet is scattered by the giant atom. We note that the photon distribution  probabilities without background state in Fig.~\ref{giantGD2}(c) are plotted by using $|(\langle\psi_G(t)|-\langle\psi_f(t)|)a^{\dag}_{k_x,k_y}|0,g\rangle|^2$ in the momentum space. We further plot $|\langle g,\vec{k}_f|S-1|\vec{k}_i,g\rangle|^2$ in Fig.~\ref{giantGD2}(d) by using above optimized coupling strengths. We find that Fig.~\ref{giantGD2}(c) agrees well with Fig.~\ref{giantGD2}(d).

To study the propagation of the symmetric target scattering single-photon wave packet, we can project the target scattering wave packet ($|\psi_G(t)\rangle-|\psi_f(t)\rangle$) in Fig.~\ref{giantGD2}(b) to four subspaces with four states denoted as $|\psi_G^{(j)}(t)\rangle$ with $j=1,2,3,4$. These four states have up-down and left-right symmetries as shown by the red-dashed lines in Fig.~\ref{giantGD2}(b). The symmetry makes sure that each state has the same photon propagating behavior. Therefore, we here focus on one state to study the propagating fidelity (${\rm PF}$) of the target scattering photons. For instance, if we consider the top-right state  $|\psi_G^{(2)}(t)\rangle$ in Fig.~\ref{giantGD2}(b), then the ${\rm PF}$ can be given as
 \begin{eqnarray}
{\rm PF}=\frac{|\langle\psi_G^{(2)}(t)|D(t)|\psi_G^{(2)}(t_0)\rangle|^2}
{|\langle\psi_G^{(2)}(t_0)|\psi_G^{(2)}(t_0)\rangle|^2},
\label{group velocity}
\end{eqnarray}
where
 \begin{eqnarray}
D(t)=\sum_{m=1}^{L_y'}\sum_{n=1}^{L_x'}a^{\dag}_{n+v_x(t-t_0),m+v_y(t-t_0)}a_{n,m}
\label{D2}
\end{eqnarray}
is the translational operation of the initial scattering single-photon wave packet with the translational distance $\vec{r}=(v_x(t-t_0),v_y(t-t_0))$.  $\vec{v}=(v_{x}, v_{y})$ is the group velocity of the target scattering wave packet. $t_0$ is the artificially selected initial time of the target scattering wave packet and
$L_x'\times L_y'$ is the size of the target scattering wave packet $|\psi_G^{(2)}(t)\rangle$. We find that the ${\rm PF}$ of the scattering photon is still close to 0.8 as shown in Fig.~\ref{giantGD2}(e)  after propagating a distance equal to its own size of $L'_x$. This behavior originates from a compact distribution of the single-photon wave packet in momentum space, as shown in Fig.~\ref{giantGD2}(c). Moreover, the dynamical evolutions of the wave packets provide us a convenient way to observe the excitation of giant atom during the scattering process. In Fig.~\ref{giantGD2}(f), we show the evolution of the excited state population $|\langle e,0|\psi_G(t)\rangle|^2$ of the giant atom.
The asymmetry for the excitation of the giant atom in Fig.~\ref{giantGD2}(f)  results in an asymmetrical photon distribution of single-photon scattering wave packet in the position space as shown in Fig.~\ref{giantGD2}(b).

Moreover, by implementing the particle swarm optimization to adjust the coupling strengths at different coupling points,  the target scattering photon state with complex configuration can also be produced. For example,  if we want to obtain the target scattering state, with photon distribution $|(\langle\psi_G(t)|-\langle\psi_f(t)|)a^{\dag}_{n,m}|0,g\rangle|^2$ without  ``background" as shown in Fig.~\ref{giantT24}(a) in position space corresponding to  the photon distribution $|(\langle\psi_G(t)|-\langle\psi_f(t)|)a^{\dag}_{k_x,k_y}|0,g\rangle|^2$ shown in Fig.~\ref{giantT24}(c) in the momentum space with multiple modes, then we can use the particle swarm optimization algorithm to obtain optimal coupling strengths, which satisfy the condition $2\lambda_{0}: \lambda_1: \lambda_2: \lambda_3: \lambda_4:  \lambda_5:\lambda_6:\lambda_7=1:-1.4332:-3.2858:-1.3093:1.1984:-1.3141:-3.2387:-0.9584$.
We also plot  $|\langle g,\vec{k}_f|S-1|\vec{k}_i,g\rangle|^2$ in Fig.~\ref{giantT24}(e) using optimized coupling strengths.  Figure~\ref{giantT24}(e) agrees well with Fig.~\ref{giantT24}(c). If we want to obtain a target scattering state with photon distribution $|(\langle\psi_G(t)|-\langle\psi_f(t)|)a^{\dag}_{n,m}|0,g\rangle|^2$ without  ``background" as shown in Fig.~\ref{giantT24}(b) in the  position space, corresponding to the photon distribution $|(\langle\psi_G(t)|-\langle\psi_f(t)|)a^{\dag}_{k_x,k_y}|0,g\rangle|^2$ in the momentum space as shown in Fig.~\ref{giantT24}(d), then we can use the particle swarm optimization algorithm to obtain another set of optimal coupling strengths, which satisfy the condition
$2\lambda_{0}: \lambda_1: \lambda_2: \lambda_3: \lambda_4:  \lambda_5:\lambda_6:\lambda_7=1:0.3509:0.2903:-0.0347:0.00684:0.1582:-0.1743:-0.9628$. We also plot  $|\langle g,\vec{k}_f|S-1|\vec{k}_i,g\rangle|^2$ in Fig.~\ref{giantT24}(f) using optimized coupling strengths.  Figure~\ref{giantT24}(f) agrees well with Fig.~\ref{giantT24}(d).
Thus, we conclude that the target scattering state can be obtained by optimizing the coupling strengths via the optimization function $Q$ in Eq.~(\ref{optimization1}) and optimization algorithms.

We note  that it is necessary to have a sufficient interaction time between the wave packet and the giant atom for realizing the scattering of single-photon scattering.
 It is clear that the self-energy function $\Sigma$ in Eq.~(\ref{self-energy})  is purely imaginary number and corresponds to the system decay with the rate $\gamma=|\Sigma|$ determined by the coupling between the resonator array and the giant atom. Thus, to realize scattering, the interaction time $\tau=L_x/v_x$ between the incident wave packet and the atom should satisfy the condition $\tau\gg1/\gamma$, where $1/\gamma$ is the lifetime of the atomic excited state. That is, the giant atom should be strongly coupled to the resonator array or the single-photon wave packet should have large enough width.
 In our numerical somulations, we assume $2J_x\tau=25$,  $2J_x/\gamma\thickapprox 0.8334$ Fig.~\ref{SMatom}(d),  $2J_x/\gamma\thickapprox4.7461$ for Fig.~\ref{giantGD2}(d), $2J_x/\gamma\thickapprox8.3091$ for Fig.~\ref{giantT24}(e), $2J_x/\gamma\thickapprox3.7397$ for Fig.~\ref{giantT24}(f). For these parameters, the atoms and the wave packets have the sufficient interaction time.

\subsection{Transmission Probabilities}

\begin{figure}[tbp]
\centering
\includegraphics[width=8cm]{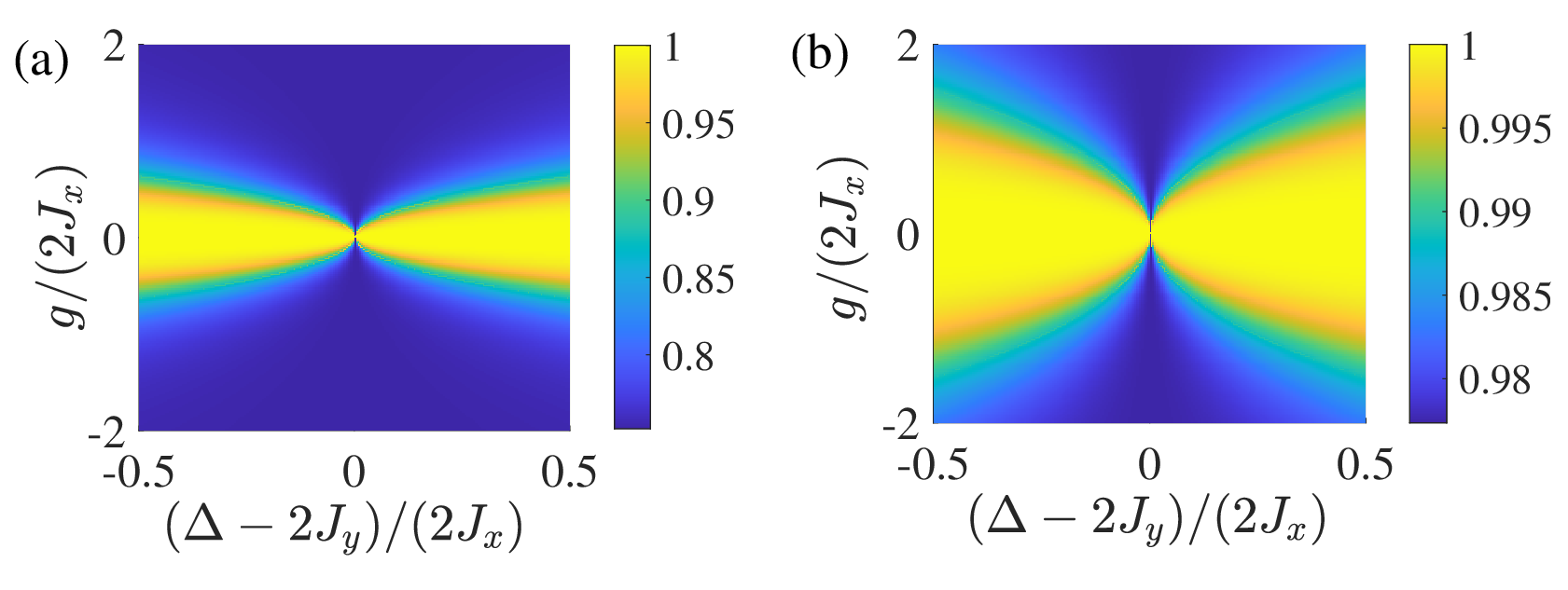}
\caption{(a) and (b) The transmission probabilities $T$ versus the coupling strength $g$ and the detuning $\Delta$ for the small atom and giant atom, respectively.
The main parameters are consistent with Fig.~\ref{giantGD2}.}
\label{TGS}
\end{figure}

We emphasize that our study mainly focuses on the distributions of the scattering photons. However,  traditional scattering problems primarily study the scattering or transmission probabilities~\cite{Zhou,Nie}. Corresponding to Fig.~\ref{SMatom} and Fig.~\ref{giantGD2}(d), we can obtain transmission probabilities $T\thickapprox0.7771$ and $T\thickapprox0.9951$, respectively, which are derived from Eq.~(\ref{TP}).  However, corresponding to Figs.~\ref{giantT24}(e) and (f),  the transmission probabilities are given as $T\thickapprox0.9749$ and $T\thickapprox0.9940$, respectively. That is, only a small portion of the photons is scattered. This is owing to the locality of the coupling patterns compared to the size of the  incident  wave packet.
In Figs.~\ref{TGS}(a) and (b), we plot the transmission probabilities $T$ versus the coupling strength $g$ and the detuning $\Delta$ for the small atom and giant atom, respectively. Here, $g$ is the coupling strength of the atom and $\Delta=\omega_l-\omega_a$.
With these parameters, there is no the Lamb shift [${\rm Re}(\Sigma)=0$]. We observe a symmetric structure with the reflectional symmetric axis $\Delta-2J_y=0$ and the transmission probabilities have the minimum values on this axis, e.g.,  $T_{\rm min}=0.7523$ for Fig.~\ref{TGS}(a) and $T_{\rm min}=0.9774$ for Fig.~\ref{TGS}(b).
When $\Delta-2J_y\neq0$, the transmission probabilities decay with the increase of $|g|$ and the transmission windows become wider as $|g|$ gets larger.

\subsection{Emission from the Giant Atom}

During the scattering process, the giant atom absorbs incoming photons and subsequently emits them.
In this subsection, we analyze the radiation dynamics of the giant atom and the spatial distribution of photons under different giant atomic configurations,  and show how the  giant atomic configuration affects atomic emission behavior. Let us assume that the system evolves to the state
$|\psi_r(t)\rangle = e^{-iHt} |\psi_i\rangle$ at time \( t \) from the initial state $|\psi_i\rangle = \sigma^+ |0, g\rangle$.
\begin{figure}[t]
\centering
\includegraphics[width=8cm]{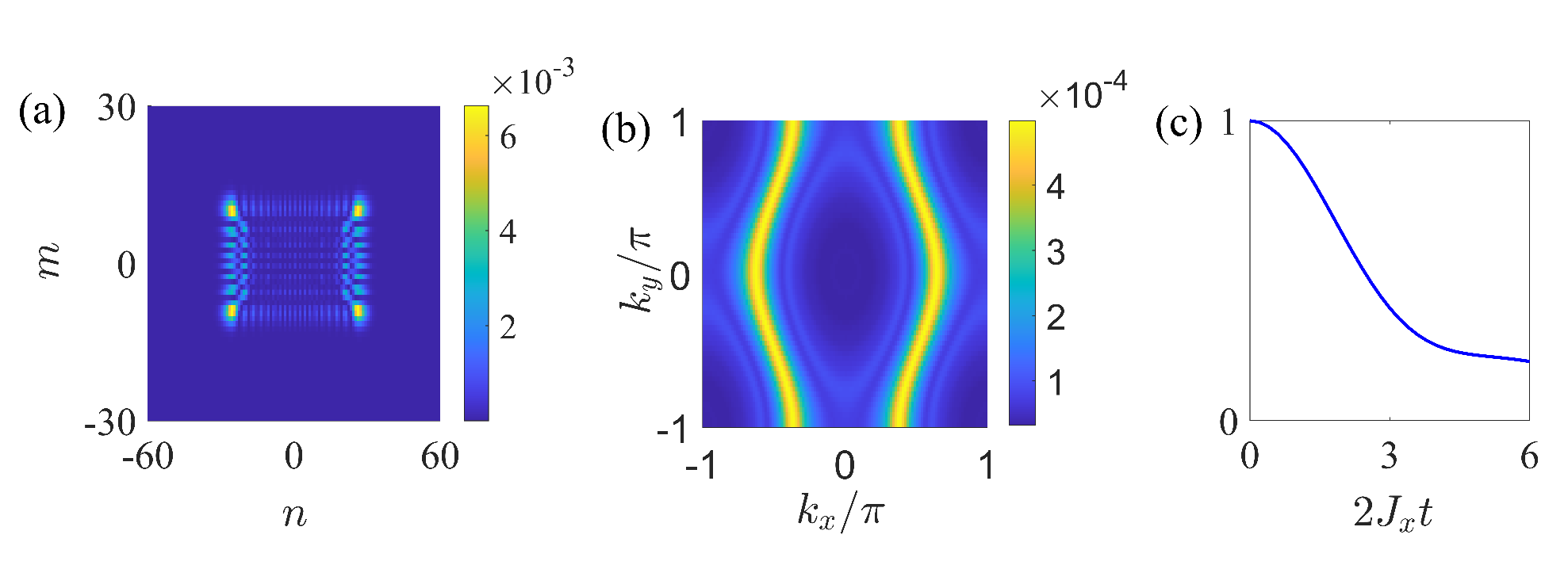}\nonumber\\
 \includegraphics[width=8cm]{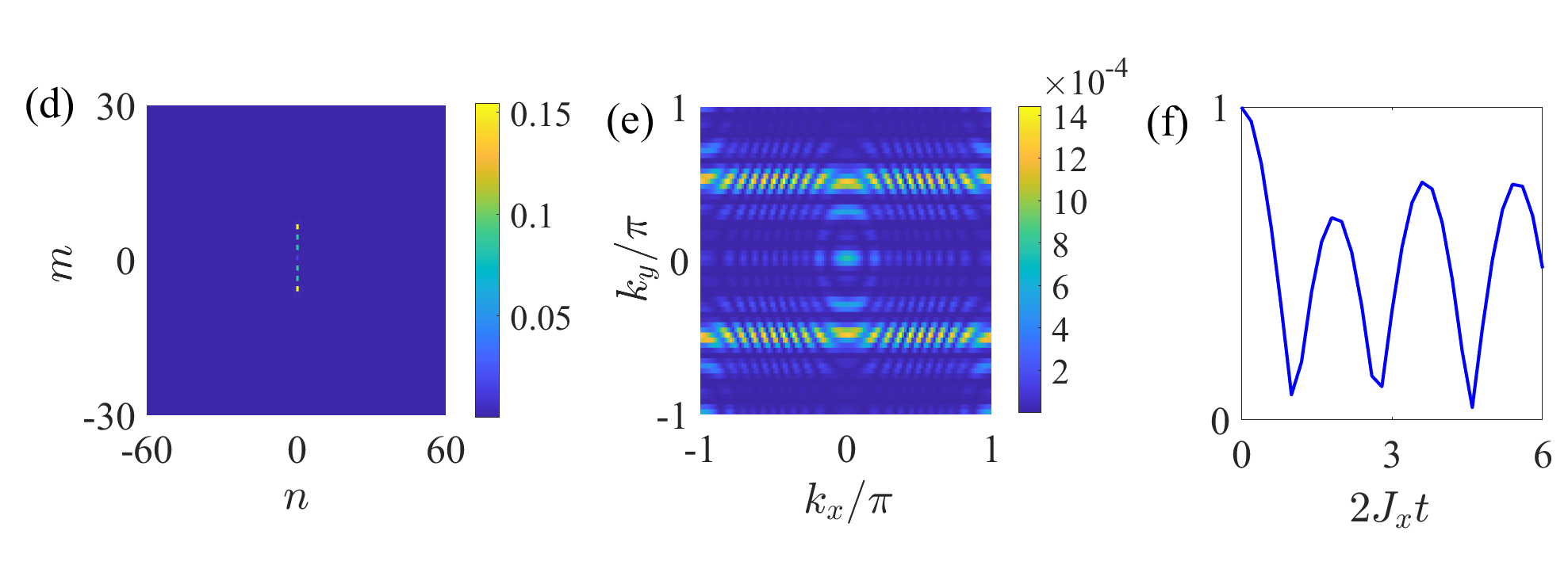}\nonumber\\
 \includegraphics[width=8cm]{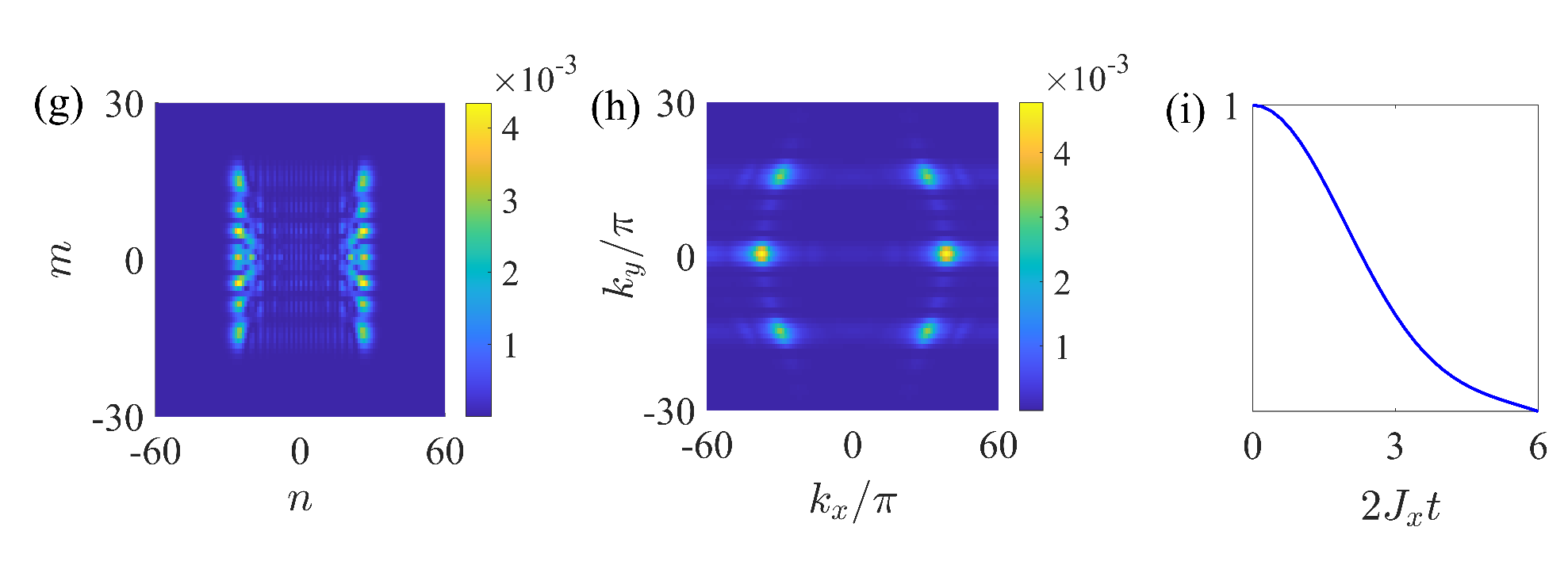}\nonumber\\
 \includegraphics[width=8cm]{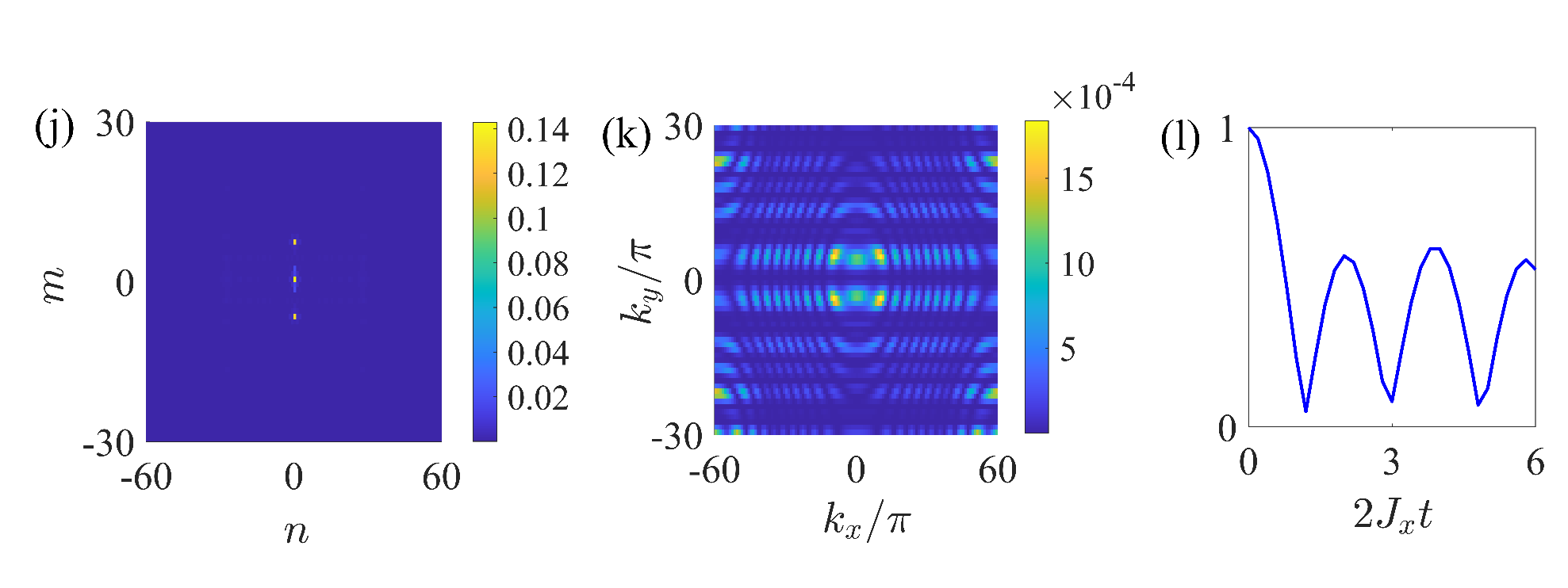}
\caption{(a), (d), (g) and (j) The distributions $|\langle \psi_r(t)|a^{\dag}_{n,m}|0,g\rangle|^2$ of emitted photons in the resonator array in the real space. (b), (e), (h) and (k) The distributions $|\langle \psi_r(t)|a^{\dag}_{k_x,k_y}|0,g\rangle|^2$ of emitted photons in the resonator array in the momentum space. (c), (f), (i) and (l) The excited state population $|\langle e,0|\psi_r(t)\rangle|^2$ of the giant atoms versus the time $t$. The parameters  for (a), (b) and (c) are the same with Fig.~(3). The parameters for (d), (e) and (f) are the same with Fig.~(4). The parameters for (g), (h), (i)  are the same with Fig.~(5)(a),  and for  (j), (k), (l)  are the same with Fig.(5) (b).}
\label{Fig1}
\end{figure}

As shown in Figs.~\ref{Fig1}(a), (b), and (c), we have plotted the photon distribution in the 2D resonator array in real space ($|\langle \psi_r(t)|a^{\dag}_{n,m}|0,g\rangle|^2$),  the momentum space ($|\langle \psi_r(t)|a^{\dag}_{k_x,k_y}|0,g\rangle|^2$), and the excited state population of  $|\langle \psi_r(t)|\sigma^+|0,g\rangle|^2$ when the 2D resonator array is coupled to a small atom via one coupling point.  Figure~\ref{Fig1}(b) shows that the emitted photons are primarily concentrated near the  constant-energy surface of the waveguide array.  Meanwhile,  Figure~\ref{Fig1}(c) shows that the dynamics of the small atom exhibits an exponential  decay.

However, we find that the photon distributions and  dynamical behaviors of the giant atom are different  when the giant atom is in the excited state and coupled to the 2D resonator array via different coupling configurations. For instance, as shown in Figs.~\ref{Fig1}(d) and (j), which have the same coupling configurations of the giant atom to the 2D resonator array with Fig.~\ref{giantGD2}(b) and Fig.~\ref{giantT24}(b), respectively, the emitted photons, characterized by $|\langle \psi_r(t)|a^{\dag}_{n,m}|0,g\rangle|^2$, mainly localize within the few lattice sites, which are coupled to the giant atom. Correspondingly, the photon $|\langle \psi_r(t)|a^{\dag}_{k_x,k_y}|0,g\rangle|^2$ in the momentum space distributes to many different constant-energy surfaces as shown in Figs.~\ref{Fig1}(e) and (k), in contrast to  Fig.~\ref{giantGD2}(d) and Fig.~\ref{giantT24}(d), respectively. In this case, the dynamics of the giant atom $|\langle \psi_r(t)|\sigma^+|0,g\rangle|^2$ exhibits oscillatory behaviors as shown in Figs.~\ref{Fig1}(f) and (l). This anomalous photon radiation behavior may be related to bound states in the continuum~\cite{BIC1,BIC2}. However, for the case shown in Fig.~\ref{Fig1}(g), which has the same coupling configuration of the giant atom to the 2D resonator array with Fig.~\ref{giantT24}(a), the emitted photons propagate in real space with nearly identical group velocity in the x-direction. This is primarily due to the photon's momentum in the x-direction being close to $k_x\sim \pm\pi/2$. In the momentum space, the emitted photons are localized near several points on the constant-energy surface as shown in Fig.~\ref{Fig1}(h).
The dynamics of the giant atom $|\langle \psi_r(t)|\sigma^+|0,g\rangle|^2$ exhibits an exponential decay as shown in Fig.~\ref{Fig1}(i).

\subsection{Scattering of a single-photon wave packet with finite size}
\label{Gaussian}

In above, we study the scattering, in which the incident single-photon wave packet covers the entire region in the y-direction, but has the finite size in the x-direction.  We now study another case that the incident single-photon wave packet has a finite distribution in both the x- and y-directions. Let us assume that an incident single-photon wave packet has two-dimensional Gaussian form
 \begin{eqnarray}
\label{eq3}
|\psi_{\rm Gauss}\rangle&=&\frac{1}{\Lambda}\sum_{n,m}e^{-(n-n_0)^2/(2\sigma_1^2)-(m-m_0)^2/(2\sigma_2^2)}\nonumber \\
&&e^{ik_ym+ik_xn}a^{\dag}_{m,n}|0,g\rangle,
 \end{eqnarray}
where $\Lambda$ is normalization constant. $n_0$ and $m_0$ are the centers of the Gaussian wave packet in the x- and y-directions, respectively. $\sigma_1$ and $\sigma_2$ are wave packet widths in the x- and y-direction, respectively. To compare with the previous results, we set $k_x=\pi/2$ and $k_y=0$, corresponding to $v_x=2J_x$ and $v_y=0$.

\begin{figure}[htbp]
\centering
  \includegraphics[width=8cm]{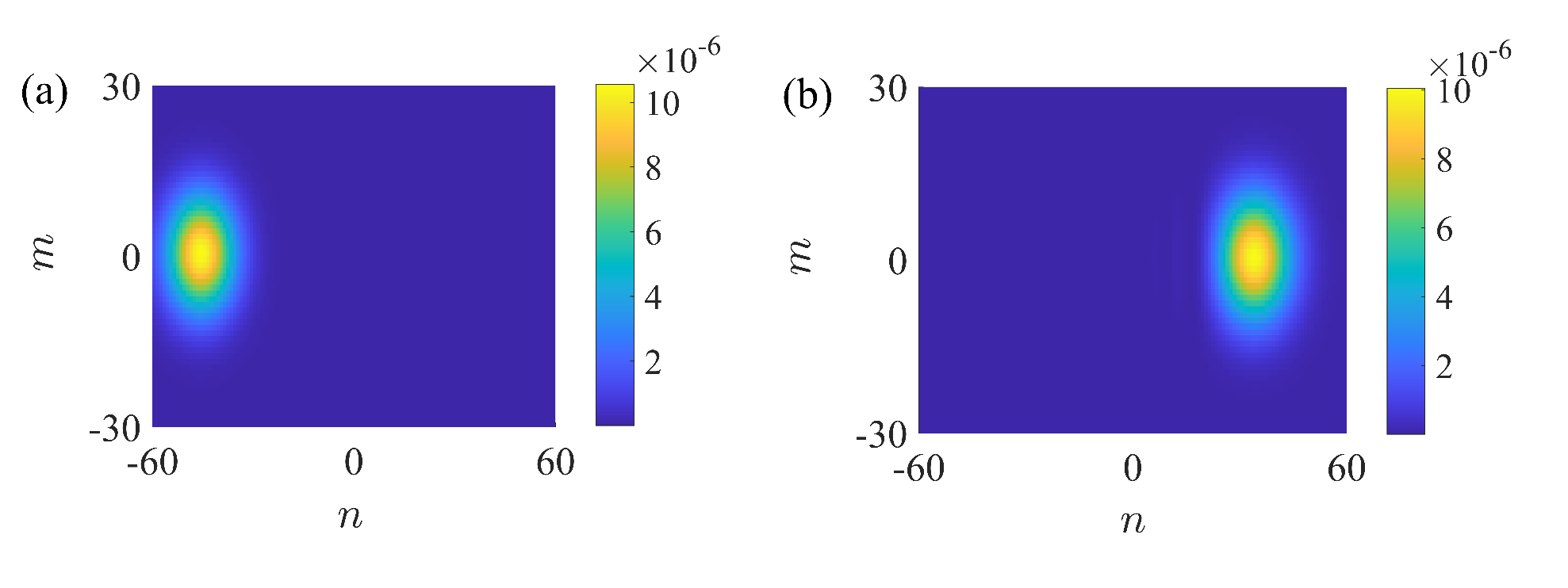}\nonumber\\
    \includegraphics[width=8cm]{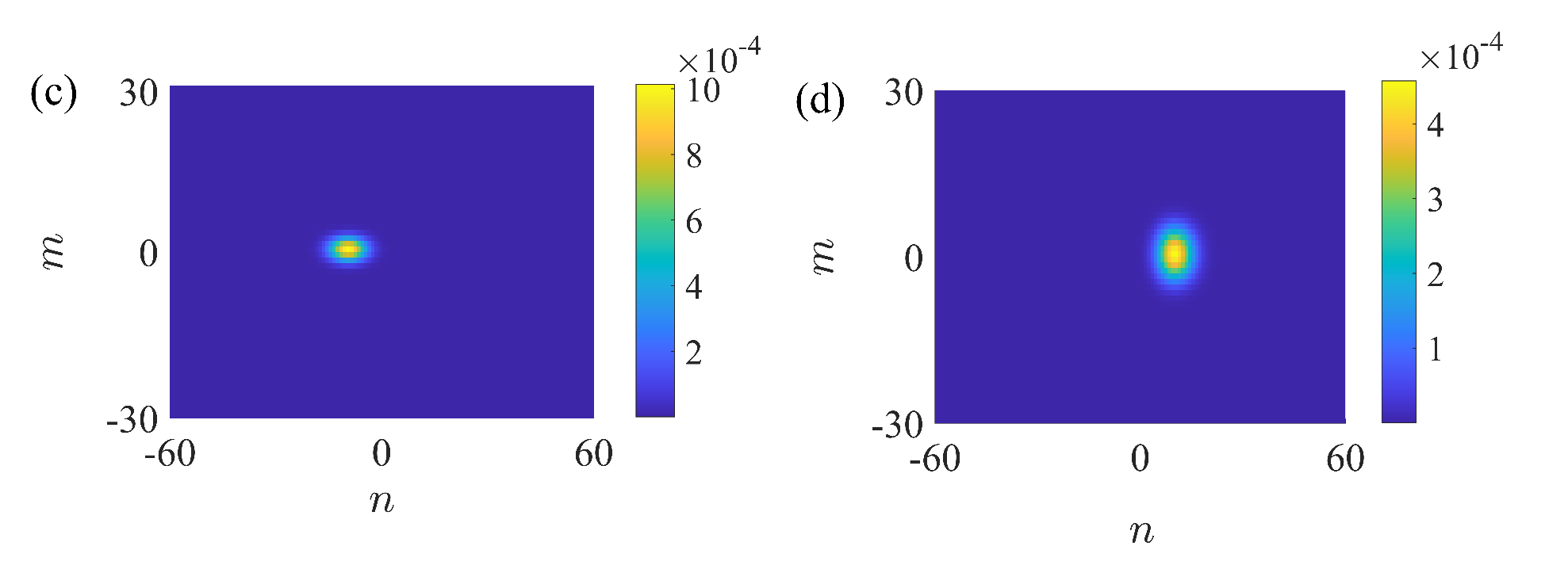}\nonumber\\
\caption{(a) and (c) The photon distributions of the Gaussian wave packets $|\langle \psi_{\rm Gauss}(t)|a^{\dag}_{n,m}|0,g\rangle|^2$ in the 2D resonator array.
(b) and (d) The photon distributions of the evolution Gaussian wave packets $|\langle \psi_{\rm f,Gauss}(t)|a^{\dag}_{n,m}|0,g\rangle|^2$ in the 2D resonator array.
The main parameters are consistent with Fig.~3 in the main text.
 The other parameters are set as: $n_0=-50$, $m_0=0$, $\sigma_1=\sigma_2=10$, $k_y=0$, $k_x=\pi/2$ and $2J_xt=80$ for (a) and (b). $n_0=-10$, $m_0=0$, $\sigma_1=5, \sigma_2=2$, $k_y=0$, $k_x=\pi/2$ and $2J_xt=20$ for (c) and (d). }
\label{Fig3}
\end{figure}

\begin{figure}[htbp]
\centering
  \includegraphics[width=8cm]{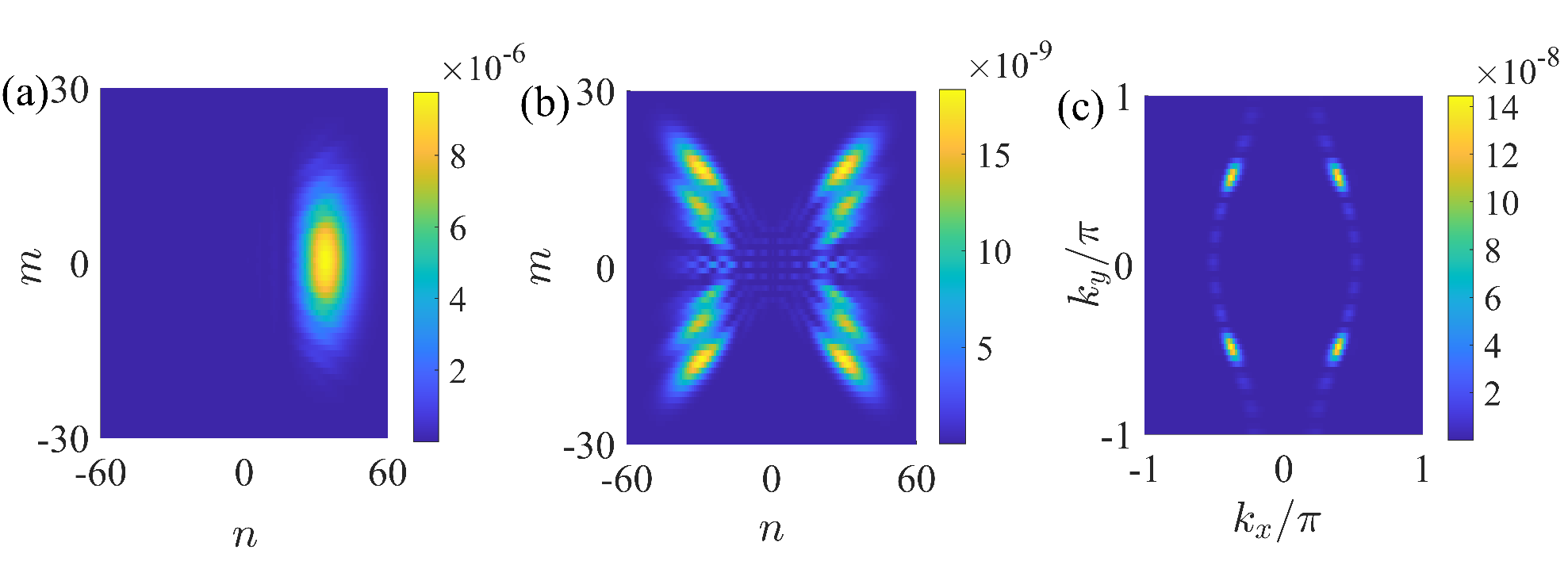}\nonumber\\
    \includegraphics[width=8cm]{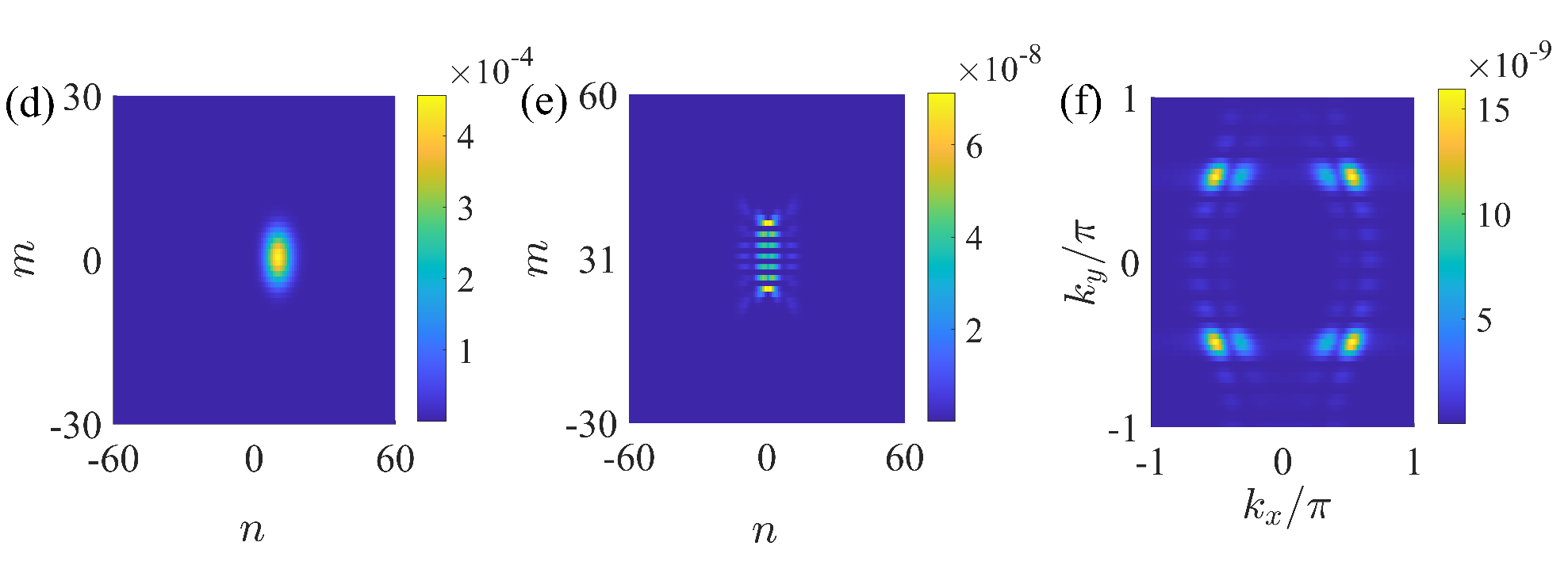}
\caption{(a), (d) and (b), (e) The photon distributions diagrams of the scattering Gaussian wave packets with $|\langle\psi_{\rm G,Gauss}(t)|a^{\dag}_{n,m}|0,g\rangle|^2$ and without $|(\langle\psi_{\rm G,Gauss}(t)|-\langle\psi_{\rm f,Gauss}(t)|)a^{\dag}_{n,m}|0,g\rangle|^2$ the ``background'' for the giant atoms, respectively.
(c) and (f) The photon distribution diagram $|(\langle\psi_{\rm G,Gauss}(t)|-\langle\psi_{\rm f,Gauss}(t)|)a^{\dag}_{k_x,k_y}|0,g\rangle|^2$ of the scattering Gaussian wave packet without the background in momentum space for the giant atom.
The main parameters are consistent with Fig.~\ref{giantGD2} and Fig.~\ref{Fig3}.}
\label{Fig4}
\end{figure}

In Figs.~\ref{Fig3}(a) and (c), we plot the photon distributions of the Gaussian wave packet for wave packet widths $\sigma_1=\sigma_2=10$ and $\sigma_1=5, \sigma_2=2$, respectively.
For Fig.~\ref{Fig3}(c), the width of the Gaussian wave packet is set to be extremely narrow, resulting in a faster diffusion speed.
Next, we study the propagating behaviors of the Gaussian wave packets in a 2D resonator array without the giant atom.
We take the two Gaussian wave packets mentioned in Figs.~\ref{Fig3}(a) and (c) as the initial states.
  With an evolution time $t$, we obtain the state $|\psi_{\rm f,Gauss}(t)\rangle=\exp(-iH_{0}t)|\psi_{\rm Gauss}(0)\rangle$ with the photon distributions $|\langle\psi_{\rm f,Gauss}(t)|)a^{\dag}_{n,m}|0,g\rangle|^2$ as shown in Figs.~\ref{Fig3}(b) and (d).
Under the parameters in Fig.~\ref{Fig3}, the Gaussian wave packet can maintain its shape well and propagate to the expected position.

Let us study the scattering effect of the Gaussian wave packet by the giant atom coupled to the 2D resonator array via 15 coupling points as for Fig.~\ref{giantGD2}.  We assume that the Gaussian wave packet is closer to the giant atom such that its shape is better maintained while passing through the giant atom.
In this case, the system's dynamics follows $|\psi_{\rm G,Gauss}(t)\rangle=\exp(-iHt)|\psi_{\rm Gauss}(0)\rangle$.
We find that the scattering effect of the giant atom on the Gaussian wave packet is very small,  the scattered photons are less than one percent of the total distribution as shown in Figs.~\ref{Fig4}(a) and (d).
After removing the background $|\psi_{\rm f,Gauss}(t)\rangle$, we can clearly observe the distribution of scattered photons in both the real space $|(\langle\psi_{\rm G,Gauss}(t)|-\langle\psi_{\rm f,Gauss}(t)|)a^{\dag}_{n,m}|0,g\rangle|^2$ and the momentum space $|(\langle\psi_{\rm G,Gauss}(t)|-\langle\psi_{\rm f,Gauss}(t)|)a^{\dag}_{k_x,k_y}|0,g\rangle|^2$, as shown in Figs.~\ref{Fig4}(b, e) and (c, f). Compared with the results shown in Fig.~\ref{giantGD2} (d), the distributions of scattered photons for numerical simulation and  analytical  expression agree well with each other. This also indicates that the wave packets can be controlled in a desired way.
However, the scattering of Gaussian wave packets by the giant atom is very weak. For the Gaussian wave packets with a width smaller than size of the giant atom, the scattered photon distribution shown in Fig.~\ref{Fig4}(f)  deviates from the analytical results shown in Fig.~\ref{giantGD2} (d). This is primarily because the wave packet does not cover all the coupling points when passing through the giant atom, and the photon intensities corresponding to each coupling point are also different.

\section{Single-Photon Scattering by a Giant Atom coupled to 2D resonator array via 2D spatial points}
\label{Chiral}

We now study how the expected asymmetric photon scattering state can be obtained for more general coupling case that the giant atom is coupled to 2D photonic resonator array via  2D spatial points not one-dimensional spatial points. We assume that all coupling points  form a 2D square lattice centered at site $(0,0)$, as described in Eq.~\eqref{HGIr}. The parameters $\lambda_{n_1,m_1}$ denote the coupling constants among different coupling points. We assume  $\lambda_{n_1,m_1}$ are complex numbers. In the following study, for  concreteness,  we assume that the giant atom is coupled to the 2D resonator array via $5\times 5$ lattice sites.

\begin{figure}[tbp]
\centering
\includegraphics[width=8cm]{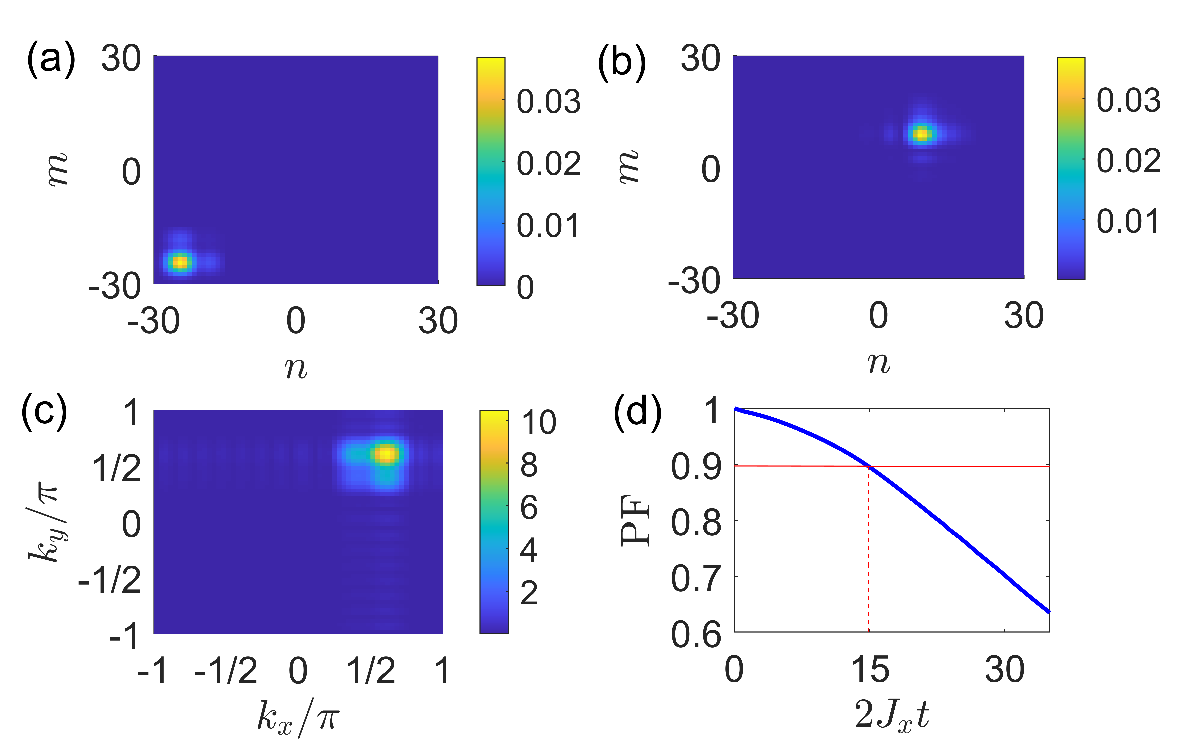}\nonumber\\
\caption{(a) The photon distribution  $|\langle \psi_l^{\prime}|a^{\dag}_{n,m}|0,g\rangle|^2$ at each lattice site $(n,m)$ for the wave packet state $|\psi_l^{\prime}\rangle$. (b) The photon distribution $|\langle \psi^{\prime}_f(t)|a^{\dag}_{n,m}|0,g\rangle|^2$ at each lattice site $(n,m)$ for the wave packet state $|\psi^{\prime}_{f}(t)\rangle$ with the evolution time $t_0=35/(2J_x)$. (c) The population distribution $|\langle\psi_l^{\prime}|a^{\dag}_{k_x,k_y}|0,g\rangle|^2$  in the moment space  for the  wave packet state $|\psi_l^{\prime}\rangle$. (d) The propagating fidelity PF of the initial wave packet versus the time $t$. The parameters are set $J_y/(2J_x)=0.5$, $N=61$, $M=61$, $L_x=15$.}
\label{2DT}
\end{figure}

Furthermore, we assume the incident single-photon propagates along the direction with a group velocity $\vec{v}\sim(2J_x,2J_y)$ in the real space. Thus, the corresponding wavevector is about $\vec{k}\sim(\pi/2,\pi/2)$ in the momentum space. We further specify the initial incident wave packet state $|\psi_l^{\prime}\rangle$ has the form:
\begin{eqnarray}
|\psi_l^{\prime}\rangle=&&\sum_{m=-(L_y/2+m_f)}^{L_y/2+m_f}\sum_{n=-(L_{x}/2)+n_{f}}^{(L_x/2)+n_{f}}\nonumber\\
&&\beta^{\prime}_{n,m}e^{ik_{c,x}n+ik_{c,y}m}a_{n,m}^{\dag}|0,g\rangle, \label{psi01}
\end{eqnarray}
which is shown in Fig.~\ref{2DT}. Where $n_f$ and $m_f$ denote the central positions of the initial wave packet in the x- and y-directions, respectively. $n$ and $m$ satisfy conditions $n_f-L_x/2\leq n\leq n_f+L_x/2$ and $m_f-L_y/2\leq m\leq m_f+L_y/2$. We find that  the free time evolution $|\psi^{\prime}_f(t)\rangle=\exp(-iH_0t)|\psi_l^{\prime}\rangle$ for the initial state $|\psi_l^{\prime}\rangle$ in the 2D resonator array  exhibits strong spatial localization after an evolution time $t$, with its central position changed to the place $(n_f+2 J_x t, m_f+2J_y t)$ where the central position $(n_f, m_f)$ of the initial wave packet reaches with the group velocity $\vec{v}\sim(2J_x,2J_y)$ and the time $t$, as shown in Fig.~\ref{2DT}(b). In the momentum space, the photon distribution is confined to be around the central wave vector $\vec{k}_{c}$ with $k_{c,x}\sim\pi/2$ and $k_{c,y}\sim \pi/2$, as shown in Fig.~\ref{2DT}(c). The compact distribution of the photon in momentum space enables the wave packet to propagate with minimal dispersion.
As shown in Fig.~\ref{2DT}(d), we plot the the ${\rm PF}$ of the wave packet versus the time $t$. The ${\rm PF}$  of the wave packet is about $90 \%$ even after propagating over a distance comparable to its own width. Below, we use the wave packet in Eq.~(\ref{psi01}) as the incident single-photon state to explore its scattering by the giant atom.

\begin{center}
\begin{table*}[htbp]
\centering
\caption{The values of $\lambda_{m1,n1}$ versus $n_1$ and $m_1$. }
\begin{tabular}{|c|c|c|c|c|c|}
\hline
    $\lambda_{m1,n1}$ & $n_1=-2$ & $n_1=-1$ & $n_1=0$ & $n_1=1$ & $n_1=2$ \\
    \hline
    $m_1=-2$ & $1$ & $-7.3486e^{3.6150i}$ & $1.862e^{0.3726i}$ & $-5.6226e^{6.4033i}$ & $-2.0644e^{8.2818i}$ \\
    \hline
    $m_1=-1$ & $8.8558e^{-1.7604i}$ & $-0.6985e^{0.8562i}$ & $-4.237e^{7.1122i}$ & $1.3402e^{-2.7336i}$ & $-7.4992e^{10.518i}$ \\
    \hline
    $m_1=0$ & $14.8622e^{9.9101i}$ & $-1.8706e^{-7.5751i}$ & $1.0688e^{4.2449i}$ & $5.8708e^{0.6388i}$ & $-0.1683e^{8.8095i}$ \\
    \hline
    $m_1=1$ & $8.0156e^{7.4858i}$ & $-5.6802e^{4.7246i}$ & $8.2794e^{10.1560i}$ & $0.1046e^{4.9715i}$ & $-2.9367e^{15.0590i}$ \\
    \hline
    $m_1=2$ & $-0.9356e^{0.6925i}$ & $4.2987e^{-2.0054i}$ & $2.7451e^{11.0834i}$ & $-0.6985e^{1.0629i}$ & $-0.20242e^{10.7395i}$ \\
\hline
\end{tabular}
  \label{table2}
\end{table*}
\end{center}

 \begin{figure}[tbp]
\centering
\includegraphics[width=8cm]{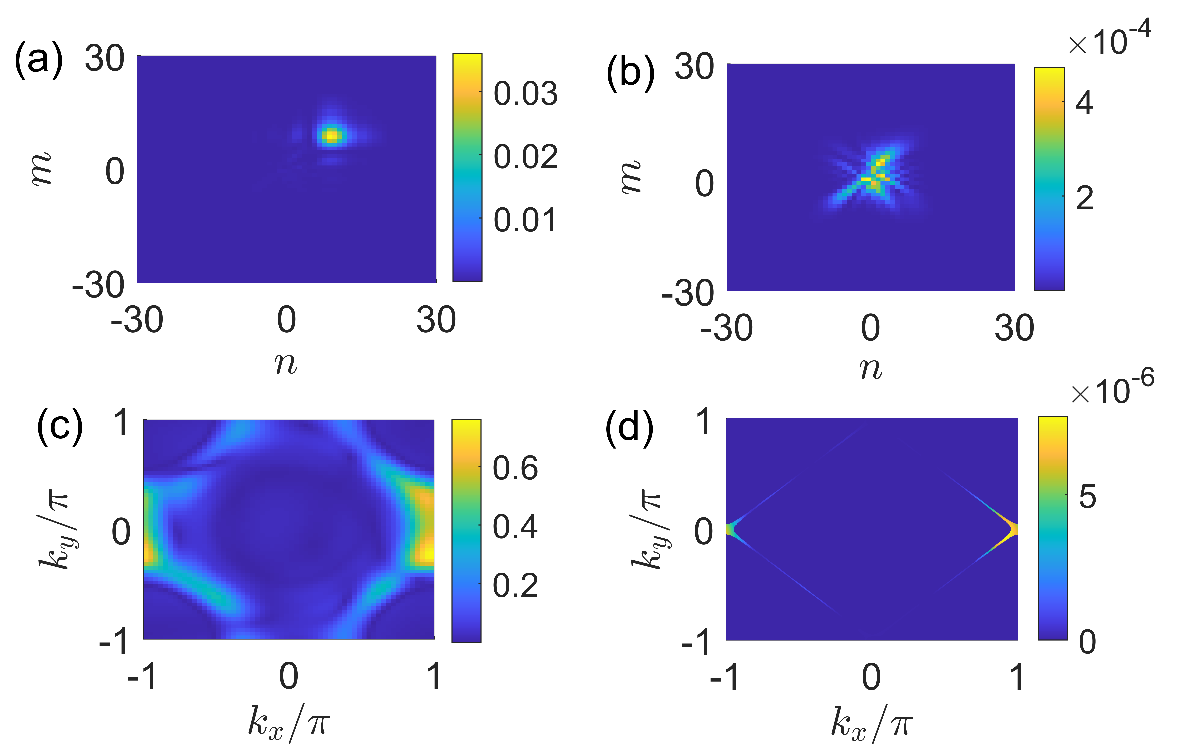}\nonumber\\
\caption{(a) and (b) The photon distributions of the scattering wave packets with $|\langle\psi^{\prime}_G(t)|a^{\dag}_{n,m}|0,g\rangle|^2$ and without $|(\langle\psi^{\prime}_G(t)|-\langle\psi^{\prime}_f(t)|)a^{\dag}_{n,m}|0,g\rangle|^2$ the ``background" for the giant atoms, respectively.
(c) The photon distribution  $|(\langle\psi^{\prime}_G(t)|-\langle\psi^{\prime}_f(t)|)a^{\dag}_{k_x,k_y}|0,g\rangle|^2$ of the scattering wave packet without the background in the momentum space. (d) The photon distribution of the photon scattering expression $|\langle g,\vec{k}_f|S-1|\vec{k}_i,g\rangle|^2$.
The main parameters are consistent with Fig.~\ref{2DT}. The other parameters are set as: $M_1=2$, $N_1=2$, $k_{ytar,1}=0$, $k_1=\pi/13$.
$t_0=35/(2J_x)$, $L_x=15$ and $L_y=15$ for (e).}
\label{2DS}
\end{figure}

We can use the optimization function Q and the particle swarm optimization (PSO) algorithm  to obtain the expected asymmetric target scattering state by optimizing the coupling strengths and phase differences among the coupling points when the initial incident state is given as in Eq.~\eqref{psi01}. For instance, we expect that the asymmetric target photon scattering state has the configuration in the momentum space as shown in Fig.~\ref{2DS}(c) after the single-photon wave packet is scattered by the giant atom with a time $t$. Then, we can  use the optimization function $Q$ and the PSO algorithm to obtain the  optimized coupling constants in Table.~\ref{table2} when the giant atom is coupled 2D resonator array via $5 \times 5$ lattice sites. For these optimized coupling constants, let us check the photon distribution of the scattered single-photon state
\begin{equation}
|\psi^{\prime}_G(t)=\exp(-iHt)|\psi_l^{\prime}\rangle
\end{equation}
 with the scattering time $t$, where $|\psi_l^{\prime}\rangle$ is given in Eq.~\eqref{psi01} and the total Hamiltonian $H$ is given in Eq.~\eqref{H} for $5\times 5$ coupling points with the optimized coupling strengths given in Table.~\ref{table2}.
In the position space,  the photon distributions with ``background" $|\langle\psi^{\prime}_G(t)|a^{\dag}_{n,m}|0,g\rangle|^2$ and without ``background" $|(\langle\psi^{\prime}_G(t)|-\langle\psi^{\prime}_f(t)|)a^{\dag}_{n,m}|0,g\rangle|^2$ are plotted in Figs.~\ref{2DS}(a) and (b), respectively.
Figure~\ref{2DS}(b) shows that the scattered photons display an asymmetric spatial distribution, with a pronounced accumulation on the right side of the giant atom.
In momentum space, the scattered photons are mainly concentrated near $\vec{k}\sim(\pi/2,0)$, as shown in Fig.~\ref{2DS}(c), consistent with the outcome of our optimization.  The group velocities $\vec{v}_x=2J_x$ and $\vec{v}_y=2J_y$ of the incident wave packet  are changed  to $v_x=2J_x$ and  $v_y=0$ after the wave packet is scattered by the giant atom.  We further plot $|\langle g,\vec{k}_f|S-1|\vec{k}_i,g\rangle|^2$ in Fig.~\ref{2DS}(d) by using above optimized coupling strengths and  it is consistent with that shown in Fig.~\ref{2DS}(c).

\section{Conclusion}\label{conclusion}

In summary, we have studied the single-photon scattering in a 2D photonic resonator array coupled to a giant atom via several spatial points with different configurations. We calculate the S-matrix of the system and propose an optimization function to control scattering photons. To simulate the scattering of single photons, we  first study the dispersion effect on the size of the single-photon wave packet during the photon propagation in 2D resonator array without coupling to the giant atom.  We  particularly show that the size of a wave packet may reach nearly stable with  the time evolution when the photon excitations of the initial wave packet aline along $y$-direction in the position space. We then consider such a wave packet with the stable size as the incident state and study the dynamical scattering of such wave packet by either a small atom or a giant atom. For the small atom, we find that the scattering photons evenly distribute around the constant energy surface of the system.
However, for the giant atom, the scattering photon distribution can be tuned by optimizing the coupling strengths between the giant atom and different lattice sites of the $2$D photonic resonator array. We show how to arbitrarily generate the  symmetric target scattering single-photon wave packet states using the giant atom. Moreover, we also study  the $\rm PF$ of the target scattering wave packets.
This is crucial for understanding the scattering process. For example, we can observe the excitation of the giant atom during scattering process that cannot be captured by the S-matrix. Moreover, we explore the controllable scattering of photons when the incident wave packet propagates along the diagonal of the lattice. By adjusting the coupling strengths at different 2D spatial connection points between the giant atom and the 2D resonator array, we induce directional interference effects, leading to an asymmetric scattered photon distribution.

 With the current technology, our model may be implementable by using superconducting quantum circuits, in which the superconducting qubits and ${\rm LC}$ resonator arrays act as the giant atoms and the microwave photonic resonator arrays, respectively~\cite{Gu-PR,Roushan,Hacohen-Gourgy,Saxberg}.
 The coupling strengths between the superconducting qubits and ${\rm LC}$ resonators can be tuned using various methods~\cite{Goren}.
The decoherence time of the superconducting qubit circuits is about $100~\mu$s, while the energy scales of  ${\rm LC}$ resonator arrays are typically in the range 100${\rm MHz}$-10${\rm GHz}$~\cite{Carroll,Georgescu}. Moreover,  the synthetic frequency dimension has been recently proposed and extensively explored in a variety of physical systems~\cite{LuqiYuan1}, e.g., in the system of the giant atoms coupled to the resonator array~\cite{Du}.  Thus,  the dynamics can be observed within the coherence time for our photon scattering proposal.  We finally point out that our proposal for the dynamical scattering of wave packets can be applied to study  the scattering problem in other systems composed of resonator arrays and atoms.  Moreover, control of propagation and scattering for light fields in $2$D photonic waveguides play a crucial role for the integrated on chip all-optical devices~\cite{Altug,Majumdar}, thus our study for the control of the scattering photons by the giant atom may have a  potential application in $2$D photonic devices.

\begin{acknowledgments}
W.C. is supported by the funding from National Science Foundation of China (Grant No. 12405017) and China Postdoctoral Science Foundation (2024M761597).
Z.W. is supported by the funding from Jilin Province (Grant Nos. 20230101357JC and 20220502002GH) and National Science Foundation of China (Grant No. 12375010).
\end{acknowledgments}

\appendix%\appendixpage
\label{app}

\addcontentsline{toc}{section}{Appendices}\markboth{APPENDICES}{}
\begin{subappendices}
\section{Derivation of the matrix elements of scattering matrix}
\label{Appendix}

According to Ref.~\cite{POGuimond}, we can calculate the matrix elements of scattering matrix in  Eq.~\eqref{Smatrix} as
\begin{widetext}
 \begin{eqnarray}
\langle g, \vec{k}_f|S|\vec{k}_i,g\rangle&=&\langle \vec{k}_f|\vec{k}_i\rangle-2\pi i\delta(E_f-E_i)\langle g, \vec{k}_f|\sum_{n=0}^{\infty}H_{\rm G,I}\left(\frac{1}{E_i-H_{0}+i0^+}H_{\rm G,I}\right)^n|\vec{k}_i,g\rangle\nonumber \\
&=&\langle \vec{k}_f|\vec{k}_i\rangle-2\pi i\delta(E_f-E_i)\langle g,\vec{k}_f| H_{\rm G,I}\sum_{n=0}^{\infty}\left(\frac{1}{E_i-H_{0}+i0^+}H_{\rm G,I}\right)^n\frac{1}{E_i-H_{0}+i0^+}H_{\rm G,I} |\vec{k}_i,g \rangle\nonumber\\
&=&\langle \vec{k}_f|\vec{k}_i\rangle-2\pi i\delta(E_f-E_i)\frac{G^{\ast}(\vec{k}_f) G(\vec{k}_i)}{E_i-\omega_a+i 0^+}\langle e,0| \sum_{n=0}^{\infty} \left(\frac{1}{E_i-H_{0}+i0^+}H_{G,I}\right)^n|0,e \rangle\nonumber \\
&=&\langle \vec{k}_f|\vec{k}_i\rangle-2\pi i\delta(E_f-E_i)\frac{G^{\ast}(\vec{k}_f) G(\vec{k}_i)}{E_i-\omega_a+i 0^+}\langle e,0|\sum_{n=0}^{\infty} \left[\frac{1}{E_i-\omega_a+i0^+}\left(\frac{H_{\rm G,I}}{E_i-H_{0}+i0^+}H_{\rm G,I}\right)\right]^n|0,e \rangle\nonumber \\
&=&\langle \vec{k}_f|\vec{k}_i\rangle-\frac{i}{2\pi}\sum_{q}\frac{\delta(\vec{k}_f-\vec{q})}{|\nabla E_f|_{\vec{k}_f=\vec{q}}}\frac{G^{\ast}(\vec{k}_f) G(\vec{k}_i)}{E_i-\omega_a-\Sigma+i0^+}.
\end{eqnarray}
\end{widetext}
The summation of the first line in Eq.~(\ref{Tmatrix}) can be rewritten to the second line by using the relation $\langle g, \vec{k}_f|H_{\rm G,I}|\vec{k}_i,g\rangle=0$. We further  insert the unit operator $|0,e\rangle\langle e,0|+\sum_{k}|k,g\rangle\langle g,k|$ into the summation of the second line, then we find that the contribution of the summation in the third line is only from the terms of  even number $n$, all terms with the odd numbers are zero. Thus, we can further rewrite the summation of the third line to the fourth line by only considering the terms of the even number. In the last line, we also use  the relation $\sum_{n=0}^{\infty} x^n=1/(1-x)$ and change the summation into the integral. From the fourth line to the fifth line, we use $\delta[\varphi(x)]=\sum_{i}\delta(x_{i})/|\varphi^{\prime}(x_{i})|$ such that $\delta(E_f-E_i)$ for energy can be changed to $\delta(\vec{k}_f-\vec{q})$ for wavevector, where $q$ is root of
the equation $E(\vec{k}_{f})-E(\vec{k}_{i})=0$ for variable $\vec{k}_{f}$ with the given initial wavevector $\vec{k}_{i}$ of the incident single-photon wavepacket.

\section{Scattering of Wave Packet by Giant Atoms with a Few Coupling Points}
\label{Appendixb}

For small atoms, scattered photons are typically distributed uniformly across the constant-energy surface in Sec.~\ref{Propagation and Scattering of Wave Packet}. However, we observe that when a giant atom is coupled to a two-dimensional resonator array via multiple coupling points, interference between these points arises, modifying the probability distribution of scattered photons on the constant-energy surface. An illustrative example is presented to demonstrate this effect.

\begin{figure}[htbp]
\centering
  \includegraphics[width=8cm]{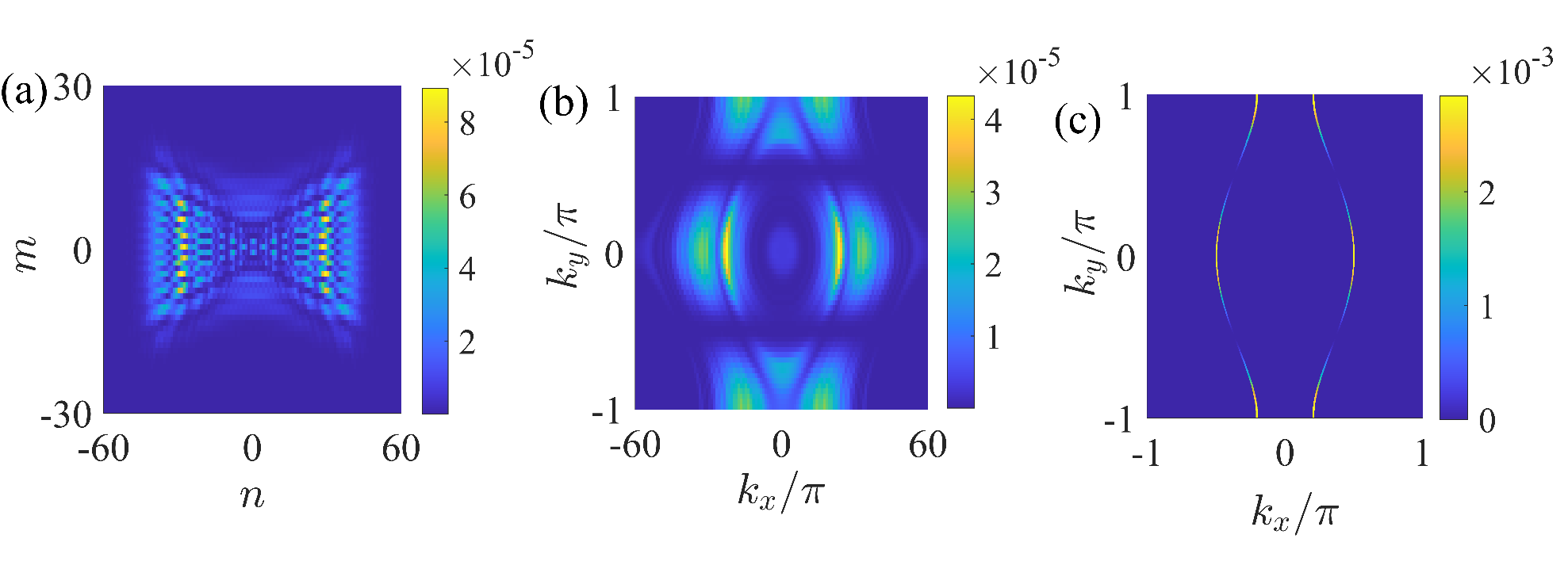}\nonumber\\
 \includegraphics[width=8cm]{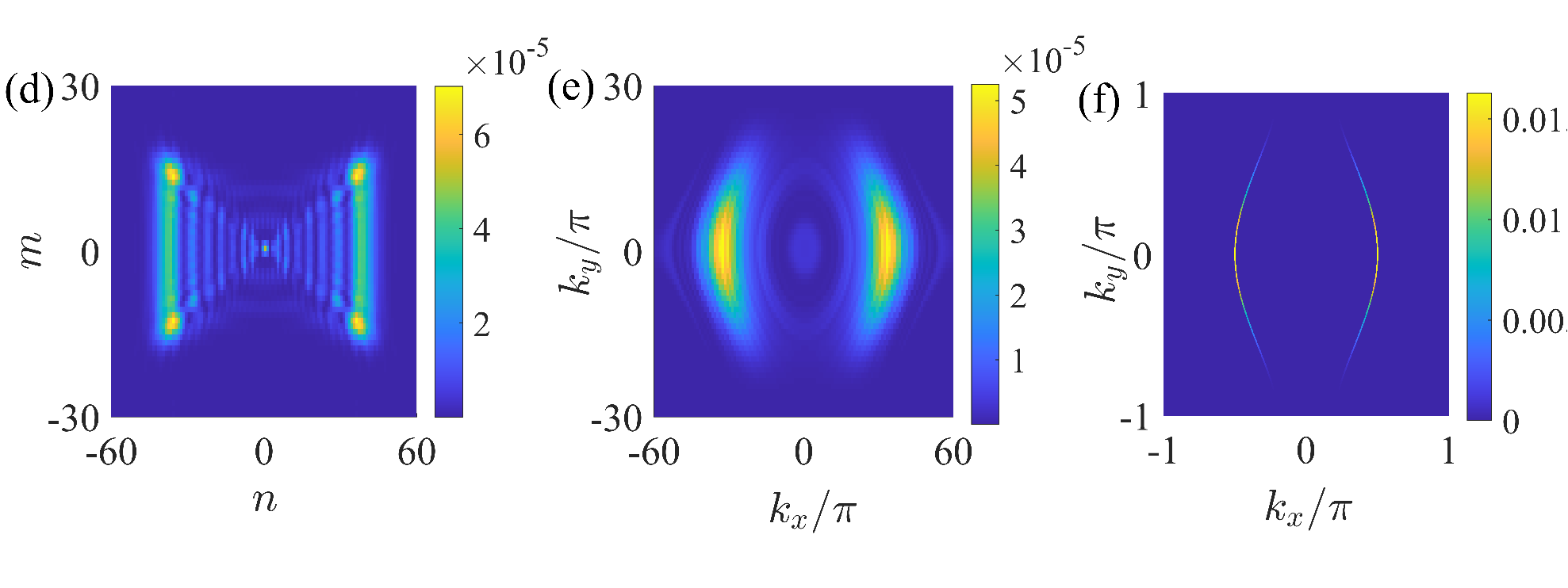}
\caption{(a) and (d) The distributions of scattering photons in the waveguide in real space for the giant atom with two and three coupling points.
(b) and (e) The distributions of scattering photons in the waveguide in momentum space for the giant atom with two and three coupling points.
(c) and (f) The distributions of the photon scattering expressions $|\langle g,\vec{k}_f|S-1|\vec{k}_i,g\rangle|^2$ for the giant atom with two and three coupling points.
The main parameters are consistent with Fig.~\ref{SMatom} in Sec.~\ref{Propagation and Scattering of Wave Packet}. For the giant atom with for two coupling points, both coupling points, having the same coupling strength $g/(2J_x)=1/2$, are located at (0,1) and (0,-1). For three coupling points, both coupling points, having the same coupling strength$g/(2J_x)=1/3$, are located at (0,1), (0,0) and (0,-1).}
\label{Fig2}
\end{figure}

In Figs.~\ref{Fig2} (a) and (d), we have plotted the distributions of scattered photons (without background) in the real space with two and three coupling points, respectively. Accordingly, the distributions of scattered photons in the momentum space are plotted in Figs.~\ref{Fig2}(b) and (e), which that the scattered photons are still distributed on the constant-energy surface. In Figs.~\ref{Fig2}(c) and (f), we have plotted scattered photon distribution of the $|\langle g,\vec{k}_f|S-1|\vec{k}_i,g\rangle|^2$ to verify  the reasonableness of Figs.~\ref{Fig2}(b) and (e).

\end{subappendices}

\end{document}